\documentclass{JHEP3}
\usepackage{axodraw}
\usepackage{amsmath}
\usepackage[utf8]{inputenc}

\usepackage{cite}
\usepackage{epsfig}
\usepackage{xspace}
\usepackage{graphics}
\usepackage{relsize}
\usepackage{amsfonts}
\usepackage{ifthen}
\usepackage[normalem]{ulem}	

\def\hide#1{}

\newcommand{\asme}{\alpha_{\textnormal{s\tiny{ME}}}}

\newcommand{\tms}{t_{\textnormal{\tiny{MS}}}}

\newcommand{\mz}{\textnormal{M}_{\textnormal{\tiny{Z}}}}
\newcommand{\mw}{\textnormal{M}_{\textnormal{\tiny{W}}}}

\newcommand{\muf}{\mu_F}
\newcommand{\mur}{\mu_R}

\newcommand{\wckkwl}[1]{w_{#1}}
\newcommand{\wumeps}[1]{w_{#1}^\prime}
\newcommand{\Oas}[1]{\mathcal{O}(\as^{#1})}
\newcommand{\Oasof}[2]{\mathcal{O}\left(\as^{#1}\left(#2\right)\right)}
\newcommand{\ordms}{\ensuremath{\rho_{\textnormal{\tiny MS}}}}

\newcommand{\dsexc}[2]{\frac{d\sigma_{#1}^{ex}}{d\phi_{#2}}}
\newcommand{\dsinc}[2]{\frac{d\sigma_{#1}^{in}}{d\phi_{#2}}}

\newcommand{\mesqof}[2]{\left|\mathcal{M}_{#1}\left(#2\right)\right|^2}

\newcommand{\Bornev}[1]{\textnormal{B}_{#1}}

\newcommand{\Iev}[2]{\int_s\widehat{\textnormal{B}}_{#1\rightarrow #2}}
\newcommand{\Tev}[1]{\widehat{\textnormal{B}}_{#1}}

\newcommand{\vincia}{V{\smaller INCIA}\xspace}

\newcommand{\rivet}{R{\smaller IVET}\xspace}

\newcommand{\pytppp}{P{\smaller YTHIA}8\xspace}

\newcommand{\as}{\ensuremath{\alpha_{\mathrm{s}}}}

\newcommand{\ECM}{\ensuremath{E_{\mathrm{CM}}}}

\newcommand{\particle}[1]{\ensuremath{\mathrm{#1}}}
\newcommand{\antiparticle}[1]{\ensuremath{\bar{\mathrm{#1}}}}

\newcommand{\g}{\particle{g}}
\newcommand{\p}{\particle{p}}

\newcommand{\W}{\particle{W}}

\newcommand{\Wm}{\ensuremath{\W^-}}
\newcommand{\Z}{\particle{Z}}

\newcommand{\qu}{\particle{u}}

\newcommand{\qc}{\particle{c}}

\newcommand{\qbar}{\antiparticle{q}}

\newcommand{\dbar}{\antiparticle{d}}

\newcommand{\cbar}{\antiparticle{c}}

\newcommand{\ord}{\ensuremath{\rho}}

\newcommand{\state}[1]{\ensuremath{S_{+#1}}}
\newcommand{\splitP}{\ensuremath{P}}

\newcommand{\done}[1]{}

\def\mrm#1{\mathrm{#1}}

\newcommand{\eg}{e.g.\xspace}

\def\sub#1{\ensuremath{_{\mrm{#1}}}}
\def\sup#1{\ensuremath{^{\mrm{#1}}}}

\def\Pnoem{\ensuremath{\Pi}}
\def\noem#1{\ensuremath{\Pnoem_{\state{#1}}}}
\def\Sud#1{\ensuremath{{\mathbf\Delta\sub{\state{#1}}}}}
\def\f2d3{\ensuremath{F_2^{\mrm{D}3}}}

\providecommand{\eqref}[1]{eq.~(\ref{#1})\xspace}
\renewcommand{\eqref}[1]{eq.~(\ref{#1})\xspace}
\newcommand{\eqsref}[1]{eqs.~(\ref{#1})\xspace}

\newcommand{\splitPDF}[1]{{\ensuremath{\mathcal{P}_{#1}}}}

\newcommand{\noemof}[2]{\ensuremath{\Pnoem_{\state{#1}}\left(#2\right)}}
\newcommand{\sudakovof}[2]{\ensuremath{\Sud{#1}\left(#2\right)}}
\newcommand{\I}[3]{\int_{#2}^{#1}\!\!\!d\rho_{#3}dz_{#3}}
\newcommand{\M}[1]{d\rho_{#1}dz_{#1}}
\newcommand{\igrand}[3]{\as\left(#2\right)\splitPDF{#1}\left(z_{#3},#2\right) }

\newcounter{aenumct}

\newcounter{ienumct}

{\end{list}}

\newcounter{enumct}
\renewenvironment{enumerate}{\begin{list}{\arabic{enumct}.}%
{\usecounter{enumct}\setlength{\topsep}{1mm}%
\setlength{\partopsep}{1mm}\setlength{\itemsep}{0mm}%
\setlength{\parsep}{1mm}}}{\end{list}}

\def\showcommentsflag{0}
\newcommand{\showcomments}{\def\showcommentsflag{1}}

\newcounter{commentcounter}%
\newcommand{\comment}[1]{\ifnum\showcommentsflag > 0%
\addtocounter{commentcounter}{1}%
\Red{\ensuremath{\ddagger^{\arabic{commentcounter}}}}%
\marginpar{\raggedright\tiny\it\Red{\ensuremath{\ddagger^{\arabic{commentcounter}}} #1}}
\fi%
}
\newcommand{\commentdel}[2]{\ifnum\showcommentsflag > 0%
\Red{\sout{#1}}\comment{#2}%
\fi
}
\newcommand{\commentadd}[2]{\ifnum\showcommentsflag > 0%
\comment{#2}\Red{#1}%
\else
#1
\fi
}
\newcommand{\commentchange}[3]{\ifnum\showcommentsflag > 0%
\Red{\sout{#2}}\comment{#3}\Red{#1}%
\else
#1
\fi
}
\newcommand{\nocomment}[1]{\ifnum\showcommentsflag > 0%
{\tiny\it\Red{\{#1}\}}
\fi%
}
\newcommand{\nocommentdel}[1]{\ifnum\showcommentsflag > 0%
\Red{\sout{#1}}%
\fi
}
\newcommand{\nocommentadd}[1]{\ifnum\showcommentsflag > 0%
\Red{#1}%
\else
#1
\fi
}
\newcommand{\nocommentchange}[2]{\ifnum\showcommentsflag > 0%
\Red{\sout{#2}}\Red{#1}%
\else
#1
\fi
}

\showcomments

\keywords{QCD, Jets, Parton Model, Phenomenological Models}
\preprint{LU-TP 12-42\\MCnet-12-14\\ arXiv:1211.4827 [hep-ph] (November 20, 2012)}

\skip\footins = 1\bigskipamount plus 2pt minus 4pt                              

\title{Unitarising Matrix Element + Parton 
    Shower merging\footnote{Work supported in parts by
    the Swedish research council (contracts 621-2009-4076 and
    621-2010-3326).}}

\author{Leif Lönnblad and Stefan Prestel\\
  Dept.~of Astronomy and Theoretical Physics, Lund University, Sweden\\
  E-mail: {\email{Leif.Lonnblad@thep.lu.se}}
    and {\email{Stefan.Prestel@thep.lu.se}}}
  
  \abstract{We revisit the CKKW-L method for merging tree-level matrix
    elements with parton showers, and amend it with an add/subtract
    scheme to minimise dependencies on the merging scale. The scheme
    is constructed to, as far as possible, recover the unitary nature
    of the underlying parton shower, so that the inclusive cross
    section is retained for each jet multiplicity separately.}

\begin{document}
 
\sloppy
 
\section{Introduction}
\label{sec:intro}


Particularly after the discovery of a Higgs-candidate resonance at
ATLAS and CMS, the precise description of multi-jet Standard Model
(SM) processes at the LHC remains crucial. Major progress has recently
been made in combining calculations of next-to-leading order (NLO)
perturbative QCD corrections with Parton Shower (PS) based Monte Carlo
event generators
\cite{Lavesson:2008ah,Hamilton:2010wh,Hoche:2010kg,Alioli:2011nr,Hamilton:2012np,Gehrmann:2012yg,Hoeche:2012yf,Frederix:2012ps}.

In light of the rapid succession of publications on merging multiple
NLO calculations with event generators
\cite{Lavesson:2008ah,Gehrmann:2012yg,Hoeche:2012yf,Frederix:2012ps},
it seems hardly arguable that this long-standing issue is
resolved. From the point of next-to-leading order accuracy, scepticism
about state-of-the-art methods is baseless. It has however also been
pointed out that sub-leading logarithmic enhancements could be left
after the merging procedure \cite{Bauer:2008qh,
  Bauer:2008qj,scet}.  It has to be stressed that
initially, this problem is not caused by the extension of tree-level
methods to NLO, but already appears for CKKW-inspired tree-level
merging schemes
\cite{Catani:2001cc,Lonnblad:2001iq,Lavesson:2005xu,Hoeche:2009rj,Lonnblad:2011xx}. The
introduction of a merging scale ($\tms$) introduces logarithmic
dependencies, $L=\ln\muf/\tms$, with a dominant contribution
$\as^nL^{2n}$, in the multi-jet tree-level configurations. These terms
are partly cancelled by the parton shower higher-order corrections to
low-multiplicity states. Let us look at one-jet merging in $\W$-boson
production, with a strictly leading-logarithmic parton
shower. Integrating the $\W$+jet matrix element (ME) over the one-jet
phase space introduces the dependencies $\as L^{2}$ and $\as L^{1}$,
while unresolved PS emissions in the zero-jet state produce $\as
L^{2}$ terms, but with negative sign. In the total cross section, the
$\as L^{2}$ terms cancel, and only a $\tms$-dependence beyond the
accuracy of the parton shower (i.e.\ $\as L^{1}$-terms) remains. This
example is rather academic, since modern parton showers include a
matrix-element correction for $\W$+jet production, thus ensuring that
in one-jet merging, all dependencies on $\tms$ are
cancelled\footnote{Note that the introduction of ME corrections
  already endows the PS with the tree-level description of the hardest
  jet, so that one-jet merging would not be necessary.}. After the
inclusion of matrix element corrections
\cite{Bengtsson:1986et,Bengtsson:1986hr,Miu:1998ut,Miu:1998ju}, the
merging scale dependence enters for two-jet merging.  However, since
control of beyond-leading logarithmic contributions cannot be
universally exerted in a parton shower, it is clear that dependencies
enter at some level, commonly hoped to be at next-to-next-to-leading
logarithmic accuracy.  The logarithmic structure of a two-jet
tree-level cross section for example has only an approximate
equivalent in parton showers, meaning that certainly at $\Oas{2}$,
sub-leading contributions are not fully cancelled. This is often, for
lack of a better term, referred to as violation of PS unitarity.

We believe it important to investigate this issue more closely. In this
publication, we take a step back from the remarkable progress in NLO merging
and outline a tree-level merging method that amends the behaviour of parton 
showers to ensure that no spurious logarithmic enhancements are introduced by
including multi-jet matrix elements. The foundation of this method is PS
unitarity, i.e.\ the requirement that the lowest order cross
section remains unchanged by methods introduced
to ameliorate the description of shapes
of observable. It should be noted that in particular the GKS 
matrix-element-correction method in \vincia \cite{Giele:2007di,Giele:2011cb} has 
emphasised a unitarity-based approach before. The aim of this article is more 
modest, in that we offer a novel prescription of combining the input used in 
multi-jet merging procedures, moving from the additive
scheme of CKKW-inspired methods to an add-and-subtract method that preserves 
the total inclusive cross section.

This article is structured as follows. Section \ref{sec:ps-unitarity}
is intended as introduction to our perspective on PS unitarity.  This
will be succeeded by a brief discussion of tree-level merging in the
CKKW-L scheme \cite{Lonnblad:2001iq,Lavesson:2005xu,Lonnblad:2011xx} and its problems in section \ref{sec:basics-ckkw-l},
before we move on to construct a LO merging method that preserves PS
unitarity in section \ref{sec:basics-umeps}. Section \ref{sec:results}
presents results for including additional jets in $\W$-production and QCD
dijet processes in the novel procedure, which we call UMEPS (Unitary
Matrix Element $+$ Parton Shower merging). Finally, we give a discussion
in section \ref{sec:discussion} and conclude in section
\ref{sec:conclusions-outlook}.

\section{Parton shower unitarity}
\label{sec:ps-unitarity}

Without any outside intervention, parton showers act on a lowest order 
seed cross section as a unitary operator. In other words, showering dresses
the constituents of a perturbatively calculated $2\to2$ process with
radiation, in order to set the stage for hadronisation. By generating
soft and collinear emissions, parton showering sums (at least) leading 
double-logarithmic enhancements to all orders.

Before detailing how this is achieved, let us introduce some notation
in order to make the formulae less cluttered. We will also classify
partons to be either resolved jets or unresolved jets at a particular
scale $\ordms$. By this, we mean that in a jet algorithm that exactly
inverts the parton shower, a parton would be resolved as a jet if the
evolution scale at which it was emitted ($\ord$), as well as jet
separations that have been changed by the emission due to recoil
effects, are above $\ordms$.  For the parton shower to be invertible
in a well-defined way, we assume the existence of on-shell
intermediate states between splittings. The index {\smaller{MS}}
foreshadows the use of this jet definition as merging scale\footnote{
  We here restrict ourselves to a particular jet (merging scale)
  definition, for the sake of clarity. We further assumed that the PS
  evolution variable is a measure of ``hardness", i.e.\ that soft and
  collinear divergences are located at $\ord\rightarrow0$. All the
  following arguments apply for a general merging scale $\tms$.}.

The parton shower approximates the 
effect of virtual corrections on observables sensitive to the scales 
$\ord_{i}$ and $\ord_{i+1}$ by integrating DGLAP splitting kernels 
$\splitP\left(z\right)$ over the unresolved phase space. This gives a term
\begin{eqnarray}
&& -\int^{\ord_{i}}_{\ord_{i+1}}
  d\ord~dz~
  \frac{\as(\rho)}{2\pi}
  \left\{
  \sum_{a\in\{ \textnormal{\tiny{outgoing}} \}}\!\sum_{j}
  \splitP_j^a\left(z\right)
~+
  \sum_{a\in\{ \textnormal{\tiny{incoming}} \}}\!\sum_{j}
  \frac{f_{j}^a(\frac{x_i^a}{z},\ord)}{f_{i}^a(x_{i}^a,\ord)}
  \splitP_j^a\left(z\right)
  \right\}\nonumber\\
&&\qquad\quad
\equiv~
 -\I{\ord_{i}}{\ord_{i+1}}{}
  \igrand{i+1}{\ord}{}~,\quad
\label{eq:ps-oj-oas-term}
\end{eqnarray}
where the first terms on the left-hand side sums all possible unresolved 
final state emissions, and the second term includes all unresolved initial 
state splittings. The ratios of parton distribution functions $f$ is only 
absent if no initial parton is taking part in the (unresolved) emissions.
The $\splitPDF{i+1}$ (and $\splitP$) notation is rather symbolic to permit a
certain degree of simplicity. We include symmetry factors and the typical 
$\ord$- and $z$-fractions from approximating the matrix element or multiplying
Jacobian factors in $\splitP$ and $\splitPDF{i+1}$, e.g. 
for an initial state splitting $\qbar\to\qbar\g$, we would have 
$\splitP = \frac{1}{\ord}\frac{1}{z}\frac{1+z^2}{1-z}$. It can be shown, by 
performing the $z$-integration for a specific splitting kernel $\splitP$, that 
the PS ``virtual corrections" in \eqref{eq:ps-oj-oas-term} indeed capture 
the leading logarithmic contributions of virtual corrections. Let us 
introduce the short-hand
\begin{eqnarray}
f_i(x_i,\ord_i) = 
f_i^+(x_i^+,\ord_i)
f_i^-(x_i^-,\ord_i)
\end{eqnarray}
The Sudakov form factor, resumming unresolved emissions between scales 
$\ord_{i}$ and $\ord_{i+1}$, is given by
\begin{eqnarray}
\sudakovof{i}{x_i,\ord_{i},\ord_{i+1}}
 &=& \frac{f_i(x_i,\ord_i)}{f_{i+1}(x_{i},\ord_{i+1})}
   \noemof{i}{x_i,\ord_{i},\ord_{i+1}}
\end{eqnarray}
where $\noemof{i}{x_i,\ord_{i},\ord_{i+1}}$ is the probability of no
emission from state $\state{i}$ between the $\ord_{i}$ and $\ord_{i+1}$. 
The no-emission probability in turn can be expressed as
\begin{eqnarray}
\label{eq:ps-noem-def}
\noemof{i}{x_i,\ord_{i},\ord_{i+1}}
 = \exp
   \left\{
   -\I{\ord_{i}}{\ord_{i+1}}{}
    \igrand{i+1}{\ord}{}
   \right\}
\end{eqnarray}
We have kept $x_i$ as an argument to remember that 
$\noemof{i}{x_i,\ord_{i},\ord_{i+1}}$ contains $x_i$-dependent PDF factors 
through $\splitPDF{i+1}$.

Let us consider the case when no PS emission above a scale $\ordms$ is
generated. The parton shower approximation of the resummed exclusive
zero-jet cross section is then
\begin{eqnarray}
\dsexc{0}{0}
 &=& f_0(x_0,\ord_0) \mesqof{0}{\ord_0,\mur} \noemof{0}{x_0,\ord_0,\ordms}\nonumber\\
\label{eq:ps-0j-exc}
 &=& f_0(x_0,\ord_0) \mesqof{0}{\ord_0,\mur}\\
 &-& f_0(x_0,\ord_0) \mesqof{0}{\ord_0,\mur}
     \I{\ord_0}{\ordms}{}\igrand{1}{\ord}{}\nonumber\\
 &+& f_0(x_0,\ord_0) \mesqof{0}{\ord_0,\mur}
     \I{\ord_0}{\ordms}{1}\igrand{1}{\ord_{1}}{1}\nonumber\\
&&\qquad\qquad\qquad\qquad\quad~\times
     \I{\ord_1}{\ordms}{2}\igrand{1}{\ord_{2}}{2}
  ~+~ \Oas{3} ~, \nonumber
\end{eqnarray}
where we have used the fact that 
\begin{eqnarray}
&& \frac{1}{2}\left( \I{\ord_0}{\ordms}{}\igrand{1}{\ord}{} \right)^2\\
&&=
 \I{\ord_0}{\ordms}{1}\igrand{1}{\ord_{1}}{1}\I{\ord_1}{\ordms}{2}\igrand{1}{\ord_{2}}{2}~.\nonumber
\end{eqnarray}
The zero-jet PS cross section is exclusive in the sense that no 
resolved emissions (i.e.\ emissions above $\ordms$) are produced. Beside 
resumming unresolved contributions to the zero-jet cross section, the parton 
shower also produces resolved emissions. The parton shower approximation to 
the cross section for
emitting the hardest jet at scale $\ord_1$ is
\begin{eqnarray}
\label{eq:ps-1j-inc}
\dsinc{1}{0}
 &=& \M{1}f_0(x_0,\ord_0) \mesqof{0}{\ord_0,\mur}\nonumber
     \igrand{1}{\ord_1}{1}
     \noemof{0}{x_0,\ord_0,\ord_1}~.
\end{eqnarray}
$\splitPDF{1}$ contains PDF ratios and kinematical factors (see definition \ref{eq:ps-oj-oas-term}).
The cross section is labelled with $in$ for ``inclusive", because the 
emission of further jets below $\ord_1$, but above $\ordms$, is still 
allowed. Note that $\dsinc{1}{0}$ is also exclusive in the sense that no 
resolved emissions above $\ord_1$ -- the scale of the \emph{first} 
emission -- are possible. In the following, we will always call a cross 
section inclusive if the parton shower can (at least in principle) produce 
further resolved emissions, and exclusive otherwise.

If only zero- and one-jet states are generated, the total cross section 
is given by the sum of exclusive zero-jet and inclusive one-jet cross 
sections.
Let us analyse the total cross section in the approximation of having 
maximally one parton shower emission. It useful to 
rewrite \eqref{eq:ps-0j-exc} with help of definition \eqref{eq:ps-noem-def}:
\begin{eqnarray}
\label{eq:ps-0j-exc-rewrite}
\dsexc{0}{0}
 &=& f_0(x_0,\ord_0) \mesqof{0}{\ord_0,\mur}\\
 &-& f_0(x_0,\ord_0) \mesqof{0}{\ord_0,\mur}
     \I{\ord_0}{\ordms}{}\igrand{1}{\ord}{}
     \noemof{0}{x,\ord_0,\ord}\nonumber
\end{eqnarray}
so that the total cross section is
\begin{eqnarray}
\label{eq:ps-inc-01}
\sigma^{in} &=& \int d\phi_0
  \left( \dsexc{0}{0} + \int \dsinc{1}{0} \right)\nonumber\\
 &=& \int d\phi_0 \Bigg(
     f_0(x_0,\ord_0) \mesqof{0}{\ord_0,\mur}\nonumber\\
 &&\quad~-~ f_0(x_0,\ord_0) \mesqof{0}{\ord_0,\mur}
     \I{\ord_0}{\ordms}{}\igrand{1}{\ord}{}
     \noemof{0}{x,\ord_0,\ord} \Bigg) \nonumber\\
 &+& \int d\phi_0
     f_0(x_0,\ord_0) \mesqof{0}{\ord_0,\mur}\nonumber
     \I{\ord_0}{\ordms}{}
     \igrand{}{\ord}{}
     \noemof{0}{x_0,\ord_0,\ord}\nonumber\\
 &=& \int d\phi_0 
     f_0(x_0,\ord_0) \mesqof{0}{\ord_0,\mur}
\end{eqnarray}
Thus, if the parton shower would stop after the first emission, the total
cross section is not changed by the application of the parton shower.

This small sketch does not really ensure that the total cross 
section is preserved after PS resummation. Parton showering usually generates
more than one emission, so that only being concerned with a single emission might not
be enough. The above argument can however be extended to any number of 
emissions. As an example, assume the PS had generated two emissions. Then, 
the one-jet cross section becomes exclusive by demanding that only one 
resolved emission has been produced,
\begin{eqnarray}
\label{eq:ps-1j-exc}
\dsexc{1}{0}
 &=& \M{1}f_0(x_0,\ord_0) \mesqof{0}{\ord_0,\mur}\nonumber\\
  && \igrand{1}{\ord_1}{1}
     \noemof{0}{x_0,\ord_0,\ord_1}
     \noemof{1}{x_1,\ord_1,\ordms}
\end{eqnarray}
and we need to add the PS approximation to the two-jet cross section
\begin{eqnarray}
\label{eq:ps-2j-inc}
\dsinc{2}{0}
 &=& \M{1}\M{2}f_0(x_0,\ord_0) \mesqof{0}{\ord_0,\mur}\nonumber\\
 &&  \times~
     \igrand{1}{\ord_1}{1}
     \noemof{0}{x_0,\ord_0,\ord_1}\nonumber\\
 &&  \times~
     \igrand{2}{\ord_2}{2}
     \noemof{1}{x_1,\ord_1,\ord_2}
     \Theta\left( \ord_1 - \ord_2 \right)
\end{eqnarray}
Now we rewrite \ref{eq:ps-1j-exc} by expanding the second no-emission 
probability
\begin{eqnarray}
\dsexc{1}{0}
 &=& \M{1}f_0(x_0,\ord_0) \mesqof{0}{\ord_0,\mur}
     \igrand{1}{\ord_1}{1}
     \noemof{0}{x_0,\ord_0,\ord_1}\nonumber\\
  &&\times~ \Bigg(
   1 -
     \I{\ord_1}{\ordms}{2}\igrand{2}{\ord_2}{2}\nonumber\\
  && \quad~ +
     \I{\ord_1}{\ordms}{2}\I{\ord_2}{\ordms}{3}\igrand{2}{\ord_{3}}{3}
   + \Oas{3}
     \Bigg)\nonumber\\
\label{eq:ps-1j-exc-expanded}
 &=& \M{1}f_0(x_0,\ord_0) \mesqof{0}{\ord_0,\mur}
     \igrand{1}{\ord_1}{1}
     \noemof{0}{x_0,\ord_0,\ord_1}\\
  &&\times~ \Bigg(
   1 -
     \I{\ord_1}{\ordms}{}\igrand{2}{\ord}{}\noemof{1}{x_1,\ord_1,\ord}
     \Bigg)\nonumber
\end{eqnarray}
where we have again used definition \ref{eq:ps-noem-def} to derive the last 
equality. 
If parton showering stops after generating maximally two emissions, the 
total cross section is given by
\begin{eqnarray}
\label{eq:ps-inc-012}
\sigma^{in} &=& \int d\phi_0
  \left( \dsexc{0}{0} + \int \dsexc{1}{0}  + \int \int\dsinc{2}{0}\right)
\end{eqnarray}
By comparing the second term in \eqref{eq:ps-1j-exc-expanded} with \eqref{eq:ps-2j-inc}, we see that any PS contribution of two resolved jets
cancels with terms containing one resolved and one unresolved jet. Already 
earlier, we saw that contributions with one resolved jet cancel against terms
with zero resolved jets. Thus, we again find  
\begin{eqnarray}
\sigma^{in} = \int d\phi_0 
     f_0(x_0,\ord_0) \mesqof{0}{\ord_0,\mur}\nonumber
\end{eqnarray}
It is easy to extend this argument to as many emissions as wanted: Whenever
the parton shower generates one emission, the change in the total 
cross section is counteracted exactly by unresolved contributions to states 
with one emission less. There is no 
need to correct the PS approximation of the zero-jet exclusive cross section
in the presence of two-jet states -- the zero-jet resummation
is oblivious of two-jet states. That parton showers are unitary is 
understandable directly from their construction, since the branching of an
underlying $n$-jet state produces a $n+1$-jet state, which overwrites 
(i.e.\ removes) the $n$-jet state in the wake of the branching.

\section{The problem with CKKW-L}
\label{sec:basics-ckkw-l}

Let us now review tree-level matrix element merging, more specifically
the CKKW-L scheme\footnote{Although most of what we discuss also
  applies to other CKKW-inspired merging schemes.}. We here focus
mainly on issues related to parton shower unitarity, and refer to
\cite{Lonnblad:2011xx} for a detailed description of CKKW-L in \pytppp
\cite{Sjostrand:2007gs} and to
\cite{Catani:2001cc,Lonnblad:2001iq,Lavesson:2005xu,Hoeche:2009rj} for
a more general introduction.

Matrix element merging procedures are designed to improve the PS description
of multi-jet observables. For this purpose, tree-level matrix element (ME)
calculations are combined with the parton shower, i.e.\ tree-level-weighted 
phase space points 
with $m$ ``hard process particles'' and $n$ additional partons
are included in the shower. In the following, we will often use the terms 
state, event, 
configuration or the symbol $\state{n}$ to refer to these $n+m-$body phase 
space points. As can be inferred from the form of $\state{n}$, we will most 
often understand that the state contains $m$ hard process particles, but not
mention these particles explicitly.

A consistent merging removes all overlap between ME states and the PS 
approximation. This is ensured by introducing a phase space cut $\ordms$ to 
separate the ME region from the PS region, and applying no-emission 
probabilities. The cut dependence is minimised by weighting configurations 
above and below $\ordms$ in identical fashion. The CKKW-L scheme constructs and 
chooses a sequence of lower-multiplicity states (a so-called parton shower 
history) for each ME event, since factors need to be generated that would, 
in the parton shower evolution, have contributed though intermediate stages.
With the help of the history, ME events will be reweighted with
\begin{eqnarray}
\wckkwl{n} &=& 
 \frac{x_0f_0(x_0,\ord_0)}{x_nf_n(x_n,\muf)}
 \times \left(\prod_{i=1}^{n}
  \frac{x_if_{i}(x_i,\ord_i)}{x_{i-1}f_{i-1}(x_{i-1},\ord_{i})}
  \right)\nonumber\\
&&\times\left(\prod_{i=1}^{n}\tfrac{\as(\ord_i)}{\as(\mur)}\right)
  \times\left(\prod_{i=1}^{n}\noem{i-1}(\ord_{i-1},\ord_i)\right)
  \times\noem{n}(\ord_n,\ordms)\label{eq:ckkwl-wgt-old}\\
&=& ~\frac{x_{n}f_{n}(x_{n},\ord_n)}{x_{n}f_{n}(x_{n},\muf)}
  ~\times~\prod_{i=1}^{n} 
  \Bigg[~\frac{\as(\ord_i)}{\asme}
     \frac{x_{i-1}f_{i-1}(x_{i-1},\ord_{i-1})}
          {x_{i-1}f_{i-1}(x_{i-1},\ord_i)}
     \noem{i-1}(x_{i-1},\ord_{i-1},\ord_i)~\Bigg]\nonumber\\
&&\times~
\noem{n}(x_n,\ord_n,\ordms)~,\label{eq:ckkwl-wgt}
\end{eqnarray}
where $\ord_{i}$ are the reconstructed splitting scales, and $\state{i}$ the
reconstructed intermediate states. The first PDF ratio in 
\eqref{eq:ckkwl-wgt-old} ensures that all ME configurations are normalised
to the same total cross section, given by the lowest order Born-level matrix 
element. The PDF ratios in brackets account for PDF factors in the shower 
splitting probabilities $\splitPDF{i}$ for initial state backward
evolution. The running of $\as$ is correctly included by the second bracket. 
Finally, double-counting is prevented by multiplying no-emission 
probabilities.

Let us investigate how the CKKW-L merging prescription changes the
lowest-order inclusive cross section. For simplicity, we will highlight
merging matrix elements with up to two additional jets with parton showers. 
In the simplest conceivable case of one-jet merging, applying CKKW-L defines 
the cross sections
\begin{eqnarray}
\label{eq:ckkwl-0j-exc}
\dsexc{0}{0}
 &=& f_0(x_0,\ord_0)
     \mesqof{0}{\ord_0,\mur} \noemof{0}{x_0,\ord_0,\ordms}\\
 &=& f_0(x_0,\ord_0) \mesqof{0}{\ord_0,\mur}
     \times
     \Bigg( 1 - \I{\ord_0}{\ordms}{}\igrand{1}{\ord}{}
  + \Oas{2} \Bigg) \nonumber\\
\dsinc{1}{0}
 &=& \M{1}f_0(x_0,\ord_0)
     \frac{\as(\ord_1)}{\as(\mur)}
     \frac{f_1(x_1,\ord_1)}{f_0(x_0,\ord_1)}
     \mesqof{1}{\ord_0,\mur}
     \noemof{0}{x_0,\ord_0,\ord_1}
\end{eqnarray}
It is crucial to note that the tree-level one-jet matrix element is in 
general different from the approximate PS splitting kernels. The inclusive 
lowest-order cross section is only preserved if 
\begin{eqnarray}
\label{eq:unitarity-condition-me1ps}
f_0(x_0,\ord_0) \mesqof{0}{\ord_0,\mur}
     \I{\ord_0}{\ordms}{}\igrand{1}{\ord}{}\\
 = \I{\ord_0}{\ordms}{1}f_0(x_0,\ord_0)
    \frac{\as(\ord_1)}{\as(\mur)}
    \frac{f_1(x_1,\ord_1)}{f_0(x_0,\ord_1)}
    \mesqof{1}{\ord_0,\mur}\nonumber~,
\end{eqnarray}
i.e.\ in the case where the first parton shower emission is distributed 
exactly according the one-jet matrix element. In this case, we would not 
have needed a merging prescription, since the PS would have already produced
the correct result.

Though correcting the first PS splitting to the full tree-level result is 
reasonably simple, correcting higher multiplicities requires significantly
more work. The \vincia program aims at solving this issue \cite{Giele:2007di,Giele:2011cb}. In general however,
we are currently forced to rely on tree-level merging to improve the descriptions of 
multi-jet observables.

If a first-splitting-corrected PS is available, unitarity violations will
enter when including matrix elements for the next higher jet multiplicity. 
Since the case of two-jet merging is also instructive for later 
considerations, we will list the contributions to the cross section below. 
\begin{eqnarray}
\label{eq:me02ps-not-expanded}
\dsexc{0}{0}
 &=& f_0(x_0,\ord_0)
     \mesqof{0}{\ord_0,\mur} \noemof{0}{x_0,\ord_0,\ordms}\\
\label{eq:me02ps-expanded}
 &=& f_0(x_0,\ord_0)
     \mesqof{0}{\ord_0,\mur}\\
 &&\times  \left(
     1 -
     \I{\ord_0}{\ordms}{}\igrand{1}{\ord}{}\noemof{0}{x_0,\ord_0,\ord}
     \right)\nonumber\\
\label{eq:me12ps-not-expanded}
\dsexc{1}{0}
 &=& \M{1}f_0(x_0,\ord_0)
     \frac{\as(\ord_1)}{\as(\mur)}
     \frac{f_1(x_1,\ord_1)}{f_0(x_0,\ord_1)}
     \mesqof{1}{\ord_0,\mur}
     \\
 &&  \noemof{0}{x_0,\ord_0,\ord_1}
     \noemof{1}{x_1,\ord_1,\ordms}\nonumber\\
\label{eq:me12ps-expanded}
 &=& \M{1}f_0(x_0,\ord_0)
     \frac{\as(\ord_1)}{\as(\mur)}
     \frac{f_1(x_1,\ord_1)}{f_0(x_0,\ord_1)}
     \mesqof{1}{\ord_0,\mur}
     \noemof{0}{x_0,\ord_0,\ord_1} \\
 &&\times  \left(
     1 -
     \I{\ord_1}{\ordms}{}\igrand{2}{\ord}{}\noemof{1}{x_1,\ord_1,\ord}
     \right)\nonumber\\
\label{eq:me22ps-not-expanded}
\dsinc{2}{0}
 &=& \M{1}\M{2}f_0(x_0,\ord_0)
     \frac{\as(\ord_1)}{\as(\mur)}
     \frac{f_1(x_1,\ord_1)}{f_0(x_0,\ord_1)}
     \frac{\as(\ord_2)}{\as(\mur)}
     \frac{f_2(x_2,\ord_2)}{f_1(x_1,\ord_2)}
     \mesqof{2}{\ord_0,\mur}
     \qquad\\
 &&  \noemof{0}{x_0,\ord_0,\ord_1}
     \noemof{1}{x_1,\ord_1,\ord_2}\nonumber
\end{eqnarray}
For a first-splitting-corrected PS all contributions not containing 
$\noem{1}$ cancel between \ref{eq:me02ps-expanded} and 
\ref{eq:me12ps-expanded}, except for the lowest order inclusive cross section.
Unitarity is then guaranteed if 
\begin{eqnarray}
&&   \I{\ord_0}{\ord_1}{} \I{\ord_1}{\ordms}{2}
     f_0(x_0,\ord_0)
     \frac{\as(\ord_1)}{\as(\mur)}
     \frac{f_1(x_1,\ord_1)}{f_0(x_0,\ord_1)}
     \mesqof{1}{\ord_0,\mur}
     \igrand{2}{\ord_2}{2}\nonumber\\
&&\qquad\qquad\qquad\qquad
     \noemof{0}{x_0,\ord_0,\ord}
     \noemof{1}{x_1,\ord,\ord_2}
     \nonumber\\
 &=& \I{\ord_0}{\ord_1}{} \I{\ord_1}{\ordms}{2}
     f_0(x_0,\ord_0)
     \frac{\as(\ord)}{\as(\mur)}
     \frac{f_1(x_1,\ord)}{f_0(x_0,\ord)}
     \frac{\as(\ord_2)}{\as(\mur)}
     \frac{f_2(x_2,\ord_2)}{f_1(x_1,\ord_2)}
     \mesqof{2}{\ord_0,\mur}\label{eq:twojet-unitarity}\\
&&\qquad\qquad\qquad\qquad
     \noemof{0}{x_0,\ord_0,\ord}
     \noemof{1}{x_1,\ord,\ord_2}\nonumber
\end{eqnarray}
For this, the splitting kernels need to exactly reproduce the matrix element, 
phase space must be fully covered by the parton shower, and the no-emission
probabilities need to be produced identically in both cases. Particularly the
requirement that the phase space is completely covered is problematic, since
parton showers commonly fill only phase space regions in which consecutive
emissions are ordered in a decreasing evolution variable.

Clearly, \eqref{eq:twojet-unitarity} is not fulfilled in standard PS
programs, which at best are correct to next-to-leading logarithmic
(NLL) accuracy. This means that the dependence on the merging scale
would vanish to order $\as^2L^4$ and $\as^2L^3$, but that there will
be a residual logarithmic dependence of order $\as^2L^2$.

In the next section, we would like to outline a method that sidesteps these
problems by using multi-jet matrix elements from the very beginning
to build the resummation for low-multiplicity states.

\section{Concepts of UMEPS}
\label{sec:basics-umeps}

The main concept we would like to emphasise is that appropriately weighted
matrix elements with additional jets can be used to induce resummation in
lower-multiplicity states.

For example, one-jet inclusive cross sections (\eqref{eq:ps-1j-inc}) can, 
by integrating over the phase space of the resolved jet, be manipulated to 
induce resummation in zero-jet cross section. No parton shower resummation 
above $\ordms$ would then be necessary in zero-jet contributions. This means
that we can reorder the parton shower formula for the inclusive cross section:
\begin{eqnarray}
\sigma^{in} &=& \int d\phi_0\left( \dsexc{0}{0} + \int \dsinc{1}{0}\right)
                \nonumber\\
            &=& \int d\phi_0
                \Bigg(  f_0(x_0,\ord_0) \mesqof{0}{\ord_0,\mur}\\
            &&\qquad  - \underbrace{f_0(x_0,\ord_0) \mesqof{0}{\ord_0,\mur}
                        \I{\ord_0}{\ordms}{}
                        \igrand{1}{\ord}{}
                        \noemof{0}{x_0,\ord_0,\ord}}_{\dsinc{1\rightarrow0}{0}}
                      + \int \dsinc{1}{0}\Bigg)\nonumber
\end{eqnarray}
and generate $\dsinc{1\rightarrow0}{0}$ \emph{explicitly from} 
$\dsinc{1}{0}$ by integrating over the emission phase space.
When including one additional jet into the parton shower, we can explicitly
preserve the inclusive cross section by adding the samples
\begin{eqnarray}
\label{eq:umeps-0j-inc}
\dsinc{0}{0}
 &=& f_0(x_0,\ord_0) \mesqof{0}{\ord_0,\mur}\\
\label{eq:umeps-1j-inc}
\dsinc{1}{0}
 &=& \M{1}f_0(x_0,\ord_0)
     \frac{\as(\ord_1)}{\as(\mur)}
     \frac{f_1(x_1,\ord_1)}{f_0(x_0,\ord_1)}
     \mesqof{1}{\ord_0,\mur} 
     \noemof{0}{x_0,\ord_0,\ord_1}\\
\label{eq:umeps-1to0}
\dsinc{1\rightarrow0}{0}
 &=& - \I{\ord_0}{\ordms}{}
     f_0(x_0,\ord_0)
     \frac{\as(\ord)}{\as(\mur)}
     \frac{f_1(x_1,\ord)}{f_0(x_0,\ord)}
     \mesqof{1}{\ord_0,\mur}
     \noemof{0}{x_0,\ord_0,\ord}
\end{eqnarray}
Before we continue, let us pause and investigate how we attach parton showers
to these samples. In zero-jet contributions, the effect of parton showers
above $\ordms$ is already included, so that we only need to start the parton 
shower at $\ordms$. If the one-jet matrix element is the highest multiplicity
sample, we allow the shower to generate emissions below $\ord_1$, as in 
traditional merging. Since cross section changes from allowing e.g.\ two 
resolved jets cancel exactly with unresolved jets in the one-jet state 
(see \eqref{eq:ps-2j-inc} and \eqref{eq:ps-1j-exc-expanded} and the 
discussion following \ref{eq:ps-inc-012}), allowing the
shower to produce resolved emissions does not invalidate unitarity.

We call this method UMEPS, for unitary matrix element + parton shower merging.
In principle, this method is as easily generalisable as traditional merging
techniques, and shows, on a more detailed level, difficulties reminiscent of
CKKW-L. To particularise, let us have a look at how two additional jets can 
be included by UMEPS. Naively, we would simply add 
\begin{eqnarray}
\label{eq:umeps-2j-inc}
\dsinc{2}{0}
 &=& \M{1}\M{2}f_0(x_0,\ord_0)
     \frac{\as(\ord_1)}{\as(\mur)}
     \frac{f_1(x_1,\ord_1)}{f_0(x_0,\ord_1)}
     \frac{\as(\ord_2)}{\as(\mur)}
     \frac{f_2(x_2,\ord_2)}{f_1(x_1,\ord_2)}
     \mesqof{2}{\ord_0,\mur}
     \qquad\quad\\
&&\qquad\qquad\times
     \noemof{0}{x_0,\ord_0,\ord_1}
     \noemof{1}{x_1,\ord_1,\ord_2}\quad\nonumber\\
\label{eq:umeps-2to1}
\dsinc{2\rightarrow1}{0}
 &=& - \M{1}\I{\ord_1}{\ordms}{}
     f_0(x_0,\ord_0)
     \frac{\as(\ord_1)}{\as(\mur)}
     \frac{f_1(x_1,\ord_1)}{f_0(x_0,\ord_1)}
     \frac{\as(\ord_2)}{\as(\mur)}
     \frac{f_2(x_2,\ord)}{f_1(x_1,\ord)}
     \mesqof{2}{\ord_0,\mur}\qquad\quad\\
&&\qquad\qquad\times
     \noemof{0}{x_0,\ord_0,\ord_1}
     \noemof{1}{x_1,\ord_1,\ord}\nonumber
\end{eqnarray}
and treat \ref{eq:umeps-2j-inc} as highest multiplicity sample. It is however
possible that due to undoing recoil effects, states with jets below $\ordms$
are produced by performing the integration in \ref{eq:umeps-2to1}. In this 
case, we take these contributions to be corrections to the zero-jet cross 
section, and integrate twice.
After this amendment, two-jet UMEPS merging contains the contributions
\begin{eqnarray}
\label{eq:umeps-2j-inc0}
\dsinc{0}{0}
 &=& f_0(x_0,\ord_0) \mesqof{0}{\ord_0,\mur}\\
\dsinc{1}{0}
 &=& \M{1}f_0(x_0,\ord_0)
     \frac{\as(\ord_1)}{\as(\mur)}
     \frac{f_1(x_1,\ord_1)}{f_0(x_0,\ord_1)}
     \mesqof{1}{\ord_0,\mur}
     \noemof{0}{x_0,\ord_0,\ord_1}\\
\label{eq:umeps-2j-rec1}
\dsinc{1\rightarrow0}{0}
 &=& - \I{\ord_0}{\ordms}{}
     f_0(x_0,\ord_0)
     \frac{\as(\ord)}{\as(\mur)}
     \frac{f_1(x_1,\ord)}{f_0(x_0,\ord)}
     \mesqof{1}{\ord_0,\mur}
     \noemof{0}{x_0,\ord_0,\ord}\\
\dsinc{2}{0}
 &=& \M{1}\M{2}f_0(x_0,\ord_0)
     \frac{\as(\ord_1)}{\as(\mur)}
     \frac{f_1(x_1,\ord_1)}{f_0(x_0,\ord_1)}
     \frac{\as(\ord_2)}{\as(\mur)}
     \frac{f_2(x_2,\ord_2)}{f_1(x_1,\ord_2)}
     \mesqof{2}{\ord_0,\mur}\qquad\quad\\
&&\qquad
     \noemof{0}{x_0,\ord_0,\ord_1}
     \noemof{1}{x_1,\ord_1,\ord_2}\nonumber\\
\label{eq:umeps-2j-rec2}
\dsinc{2\rightarrow1}{0}
 &=& - \M{1} \I{\ord_1}{\ordms}{}
     f_0(x_0,\ord_0)
     \frac{\as(\ord_1)}{\as(\mur)}
     \frac{f_1(x_1,\ord_1)}{f_0(x_0,\ord_1)}
     \frac{\as(\ord)}{\as(\mur)}
     \frac{f_2(x_2,\ord)}{f_1(x_1,\ord)}
     \mesqof{2}{\ord_0,\mur}\nonumber\\
  &&\qquad   \Theta(\ord_1 - \ordms)
     \noemof{0}{x_0,\ord_0,\ord_1}
     \noemof{1}{x_1,\ord_1,\ord}\\
\dsinc{2\rightarrow0}{0}
 &=& - \I{\ord_0}{\ord_1}{a}\I{\ord_1}{\ordms}{b}
     f_0(x_0,\ord_0)
     \frac{\as(\ord_a)}{\as(\mur)}
     \frac{f_1(x_1,\ord_a)}{f_0(x_0,\ord_a)}
     \frac{\as(\ord_b)}{\as(\mur)}
     \frac{f_2(x_2,\ord_b)}{f_1(x_1,\ord_b)}
 \mesqof{2}{\ord_0,\mur}\nonumber\\
  &&\qquad   \Theta(\ordms - \ord_a)
     \noemof{0}{x_0,\ord_0,\ord_a}
     \noemof{1}{x_1,\ord_a,\ord_b}\label{eq:umeps-2j-rec22}
\end{eqnarray}
UMEPS
can then be extended to arbitrary jet multiplicity. The main idea is
that in order keep unitarity, we have to subtract all contributions
that we add as higher multiplicity matrix elements. The subtractions
are constructed with PS unitarity as a guideline. For brevity, we
introduce the short-hands
\begin{eqnarray*}
\dsinc{n}{0} ~=~ \Bornev{n}\wumeps{n} ~=~ \Tev{n} \qquad\textnormal{and}\qquad
\dsinc{n\rightarrow m}{0} ~=~ -\int d^{n-m}\phi~\Bornev{n}\wumeps{n}
 ~=~ - \Iev{n}{m}
\end{eqnarray*}
where $\wumeps{n}$ will be defined below. The symbol $\Iev{n}{m}$ indicates 
that more than one integrations had to be performed since all of the states 
$\state{n-1},\dots,\state{m+1}$ contained partons below the merging scale. The
integration(s) will be achieved by substituting the input event with a 
reconstructed lower-multiplicity event of the parton shower history, as
will be discussed in section \ref{sec:umeps-step-by-step}. This substitution
method is indicated by the subscript $s$ on the integral sign. The weight 
$\wumeps{n}$ that needs to be applied to tree-level events to produce the
$\dsinc{n}{}$ sample is given by
\begin{eqnarray}
\wumeps{n}
&=&  \frac{x_{n}f_{n}(x_{n},\ord_n)}{x_{n}f_{n}(x_{n},\muf)}
~\times~
  \prod_{i=1}^{n} \Bigg[~\frac{\as(\ord_i)}{\as(\mur)}
     \frac{x_{i-1}f_{i-1}(x_{i-1},\ord_{i-1})}
          {x_{i-1}f_{i-1}(x_{i-1},\ord_i)}
     ~\noem{i-1}(x_{i-1},\ord_{i-1},\ord_i)~\Bigg]
   \qquad\label{eq:umeps-wgt}\\
&=&  \tfrac{x_{n}^+f_{n}^+(x_{n}^+,\ord_n)}{x_{n}^+f_{n}^+(x_{n}^+,\muf)}
     \tfrac{x_{n}^-f_{n}^-(x_{n}^-,\ord_n)}{x_{n}^-f_{n}^-(x_{n}^-,\muf)}
   \nonumber\\
&&\qquad\times
  \prod_{i=1}^{n} \Bigg[~\!\tfrac{\as(\ord_i)}{\as(\mur)}
     \tfrac{x_{i-1}^+f_{i-1}^+(x_{i-1}^+,\ord_{i-1})}
          {x_{i-1}^+f_{i-1}^+(x_{i-1}^+,\ord_i)}
     \tfrac{x_{i-1}^-f_{i-1}^-(x_{i-1}^-,\ord_{i-1})}
          {x_{i-1}^-f_{i-1}^-(x_{i-1}^-,\ord_i)}~\noem{i-1}(x_{i-1},\ord_{i-1},\ord_i)~\!\Bigg]
  ~.\nonumber
\end{eqnarray}
This weight differs from the CKKW-L weight in \eqref{eq:ckkwl-wgt}, since it 
does not contain the last no-emission probability 
$\noem{n}(x_n,\ord_n,\ordms)$, i.e.\ the last line in \eqref{eq:ckkwl-wgt}. In
the UMEPS procedure, this factor is instead included by subtracting the 
integrated, reweighted, next-higher multiplicity sample, thus conserving 
unitarity in a way reminiscent of standard parton showers. The probability of
having no resolved emissions off the zero-jet states in 
\eqref{eq:umeps-2j-inc0} for example, is included through the contributions in  
\eqsref{eq:umeps-2j-rec1} and \eqref{eq:umeps-2j-rec22}.

Armed with this notation, the prediction of an 
observable $\mathcal{O}$ in $2-$jet merged UMEPS becomes 
\begin{eqnarray}
  \langle \mathcal{O} \rangle
  &=& \int d\phi_0\Bigg\{ \mathcal{O}({\state{0j}}) \Tev{0}
  ~-~ \mathcal{O}({\state{0j}})\Iev{1}{0}
  ~-~ \mathcal{O}({\state{0j}})\Iev{2}{0}\nonumber\\
  &&\qquad\quad
  + \int \mathcal{O}({\state{1j}})\Tev{1}
  ~-~ \int\mathcal{O}({\state{1j}})\Iev{2}{1}\nonumber\\
  &&\qquad\quad
  + \int\!\!\!\int\mathcal{O}({\state{2j}})\Tev{2} \Bigg\}~,
\end{eqnarray}
where we have used the notation $\state{nj}$ to indicate states with $n$ 
resolved jets, resolved meaning above the cut $\ordms$ as defined by the 
merging scale definition. More generally, the outcome of merging $n$ 
additional partons with the UMEPS method is
\begin{eqnarray}
\langle \mathcal{O} \rangle
 &=& \int d\phi_0\Bigg\{ \mathcal{O}({\state{0j}}) \left[ \Tev{0}
 ~-~ \Iev{1}{0}
 ~-~ \Iev{2}{0}
 ~-~
\dots
 ~-~ \Iev{N}{0}~\right]\nonumber\\
&&\qquad\quad
 + \int \mathcal{O}({\state{1j}})\left[\Tev{1}
 ~-~ \Iev{2}{1}
 ~-~
\dots
 ~-~ \Iev{N}{1}~\right]\nonumber\\
&&\qquad\quad
 +\phantom{\int} \dots\nonumber\\
&&\qquad\quad
 + \int\!\dots\!\int \mathcal{O}({\state{N-1j}}) \left[\Tev{N-1}
 ~-~ \Iev{N}{N-1} ~\right]
\nonumber\\
&&\qquad\quad
 + \int\!\dots..\!\int \mathcal{O}({\state{Nj}})~\Tev{N} ~\Bigg\}\nonumber\\
&=&
  \sum_{n=0}^N \int d\phi_0 
  \int\!\dots\!\int
  \mathcal{O}({\state{nj}})~\left\{
  \Tev{n}
 ~-~ 
  \sum_{i=n+1}^{N}
  \Iev{i}{n}
  ~\right\}
~.
  \label{eq:UMEPS-master}
\end{eqnarray}
The generation of $\Tev{n}-$ and $\Iev{n}{m}-$events will be
summarised in section \ref{sec:umeps-step-by-step}.  It should be
noted that the treatment $\ordms-$unordered integration results is
heavily influenced by how CKKW-L includes states with
$\ordms-$unordered emissions, which was discussed in detail in
\cite{Lonnblad:2011xx}. Precisely for states which evolve from a state
below $\ordms$ to a state above $\ordms$ do CKKW-L and the truncated-shower
\cite{Nason:2004rx} approach differ. It can thus be imagined that
other ways of treating such notorious configurations show improved
behaviour. For now, we will not discuss such possibilities, and
instead, if necessary, integrate multiple times, until a state above
$\ordms$ is produced.

\FIGURE{
\centering
  \includegraphics[width=0.6\textwidth]{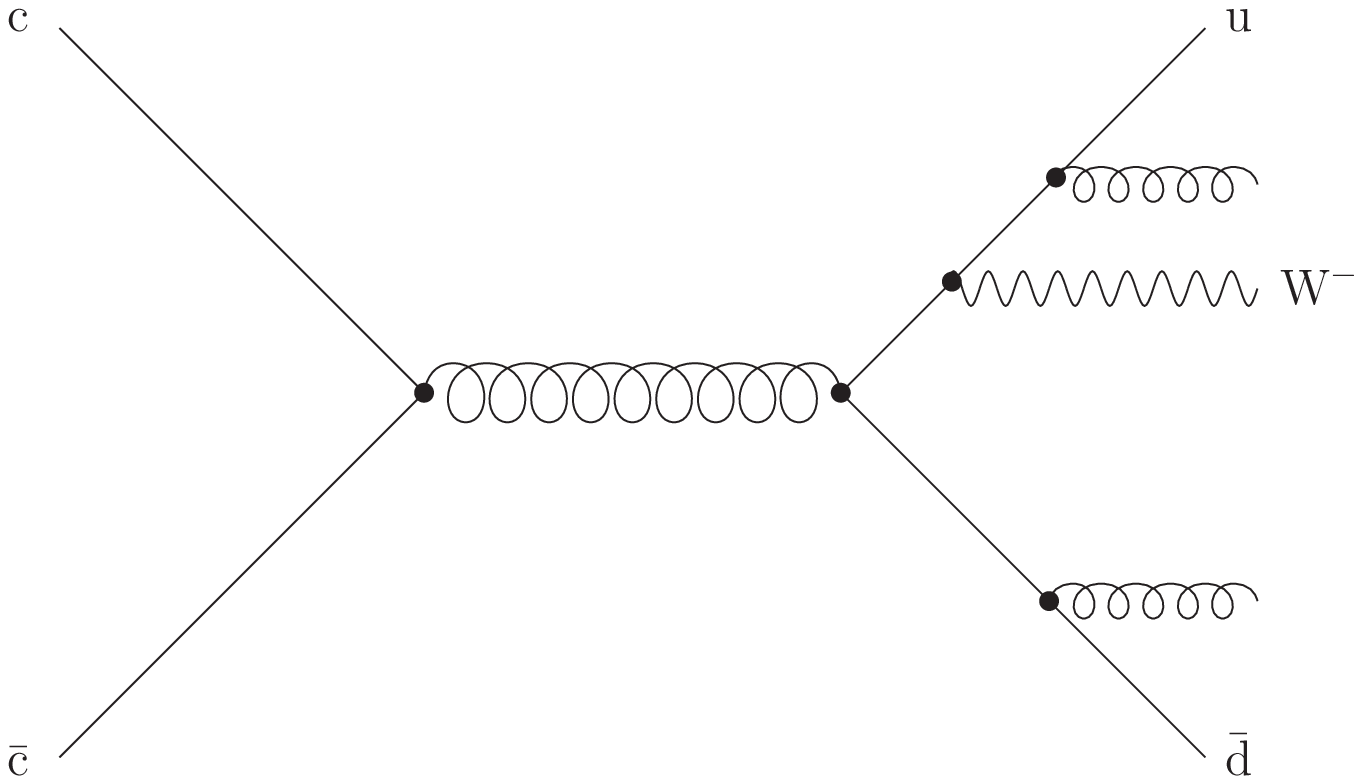}
\caption{\label{fig:incomplete}An example of a matrix element 
contribution without a complete shower history. If the parton shower does not
include $\W-$boson radiation, only the two gluon emissions can be 
reclustered, and $\qc\cbar\to\qu\dbar\Wm$ has to be considered a separate 
hard process.}
}

A well-known challenge of merging prescriptions is the treatment of
configurations that could never have been produced by a sequence of
shower splittings. This can happen if the PS does not include all
possible splittings of the model. Figure~\ref{fig:incomplete} for
example cannot be produced by a shower that does not allow $\W-$boson
radiation. Such states cannot be projected onto a lower-multiplicity
underlying process, and will thus introduce small unitarity violations
in UMEPS.

\subsection{Procedure step-by-step}
\label{sec:umeps-step-by-step}

To implement UMEPS, we need to be able to perform the necessary integrations.
Although the formulae could convey a feeling of complexity, these integrations
are factually already needed in traditional merging approaches. All modern
CKKW-inspired schemes need to construct a history of parton shower states 
for input matrix element events, because otherwise, no trial showers can be 
used to generate Sudakov form factors dynamically. Thus, a sequence
of states $\state{n}\rightarrow\state{n-1}\rightarrow\dots\rightarrow\state{1}
\rightarrow\state{0}$ is always available. The sequence is 
constructed by inverting the shower mapping of radiative phase space on each
state, i.e.\ a parton in $\state{n}$ is removed, and its momentum distributed
amongst the remaining particles, leading to a state $\state{n-1}$\footnote{
The mapping used for the current paper is given in Appendix B.2 of 
\cite{Lonnblad:2011xx}}. This is exactly the integration we need: To produce 
the integrated version of $\state{n}$, we simply replace it by $\state{n-1}$, 
but keep the full weight. $k-$fold integrations can be achieved by replacing 
$\state{n}$ with $\state{n-k}$.

With this, we have all ingredients to construct the UMEPS scheme. If not 
mentioned explicitly, all weights in UMEPS are generated precisely as in 
CKKW-L. To avoid unnecessary complications, we will here gloss over many 
technicalities that have already been addressed in CKKW-L, and are directly 
borrowed by UMEPS. A discussion of Sudakov reweighting for states without 
parton-shower ordered histories, for 
example, can be found in \cite{Lonnblad:2011xx}. 

The UMEPS algorithm has two parts -- a part in which we keep the 
matrix element configurations ($\Tev{n}-$events) and another in which 
we integrate over emissions ($\Iev{n}{m}-$events):
\begin{enumerate}
\item[I.] Produce \textit{Les Houches} event files (LHEF)
  \cite{Alwall:2006yp} with a matrix element generator for
  $n~=~0,1\ldots N$ extra jets with a regularisation cut-off, $\ordms$,
  typically using a fixed factorisation scale, $\muf$, and a fixed
  $\as(\mur)$.
\item[II.] Pick a jet multiplicity, $n$, and a state $\state{n}$ according
  to the cross sections given by the matrix element generator.
  \begin{enumerate}
  \item[1.] Find all shower histories for the state $\state{n}$, pick a sequence 
    according to the product of splitting probabilities. Only pick un-ordered 
    sequences if no ordered sequence was found. Only pick incomplete paths if no
    complete path was constructed.
  \item[2.] Perform reweighting: For each
    $0\leqslant i<n$,
    \begin{enumerate}
    \item[i.] Start the shower off the state $\state{i}$ at
      $\ord_{i}$, generate a trial state $R$ with scale
      $\ord_{R}$. If $\ord_{R} > \ord_{i+1}$, veto the event and
      start again from II.
    \item[ii.] Calculate the weight factor
      \begin{equation}
        w_{i} = \frac{\as(\ord_{i+1})}{\as(\mur)}
        \frac{x_{i}^+f_{i}^+(x_{i}^+,\ord_{i} )}
              {x_{i}^+f_{i}^+(x_{i}^+,\ord_{i+1} )}
        \frac{x_{i}^-f_{i}^-(x_{i}^-,\ord_{i} )}
             {x_{i}^-f_{i}^-(x_{i}^-,\ord_{i+1} )}
      \end{equation}
    \end{enumerate}
  \item[3.] Start the shower from $\state{n}$.
    \begin{enumerate}
    \item[i.] If $n<N$, start the shower at $\ord_n$, veto any shower emission
              producing an additional resolved jet.
    \item[ii.] If $n=N$, start the shower at $\ord_n$.
    \end{enumerate}
  \end{enumerate}
\item[III.] If the event was not rejected, multiply the event weight by
  \begin{equation}
    \wumeps{n} =
    \frac{x_{n}^+f_{n}^+(x_{n}^+,\ord_{n} )} {x_{n}^+f_{n}^+(x_{n}^+,\muf )}
    \times
    \frac{x_{n}^-f_{n}^-(x_{n}^-,\ord_{n} )} {x_{n}^-f_{n}^-(x_{n}^-,\muf )}
    \times\prod_{i=0}^{n-1} w_{i}
  \end{equation}
\item[V.] Start again from II.
\end{enumerate}

\noindent
The second part, i.e.\ producing $\Iev{n}{m}-$events to effect 
lower-multiplicity PS resummation, requires only two changes:
\begin{enumerate}
\item[II.3] Replace the matrix-element state by $\state{n-1}$, or 
            the first state $\state{l}$ with all $l\leq n-1$ partons above the
            merging scale. If no integrated state can be constructed, i.e.\ if
            only incomplete paths were found, reject the event. For valid 
            events, start the shower at $\ord_n$, veto any shower emission
            producing an additional resolved jet.
\item[III.] If the event was not rejected, multiply the event weight by
  \begin{equation}
    -\wumeps{n} =
    -\frac{x_{n}^+f_{n}^+(x_{n}^+,\ord_{n} )} {x_{n}^+f_{n}^+(x_{n}^+,\muf )}
    \times
    \frac{x_{n}^-f_{n}^-(x_{n}^-,\ord_{n} )} {x_{n}^-f_{n}^-(x_{n}^-,\muf )}
    \times\prod_{i=0}^{n-1} w_{i}
  \end{equation}
\end{enumerate}

\noindent
Finally, all samples generated in the first and second parts are added to 
give the UMEPS prediction. Note that in order produce correctly normalised 
cross sections $\dsexc{i}{0}$, we need to include ratios of parton 
distributions and $\as$ ratios into the weight. This is analogous to the 
CKKW-L method (see the $\as$- and PDF-factors in \eqref{eq:ckkwl-wgt}). It is 
worthwhile to notice that the UMEPS scheme can in principle be implemented 
by using structures already existing in traditional merging codes. 
Basically, compared to traditional merging, the $\Tev{n}-$contributions
do not carry a no-emission probability for emissions off the ME event.
The $\Iev{n}{m}-$samples can easily be extracted from merging codes.

How multiple partonic interactions (MPI) are added to the merged
samples requires a short discussion. In principle, we stay true to the
philosophy of the algorithm outlined in \cite{Lonnblad:2011xx}, i.e.\
we want to make sure that the merging method does not artificially
suppress hard secondary scatterings, which in \pytppp are interleaved
with the parton shower. The interleaving means that the PS is
competing with the MPI's, and the probability of emitting a parton in
the PS is not only governed by the standard no-emission probability,
$\noem{n}$ but is also multiplied by a no-MPI probability,
$\noem{n}\sup{MPI}$. Hence all $m$-jet (both $\Tev{m}$ and
$\Iev{n}{m}$) samples generated by our algorithm above need to be
multiplied by the no-MPI probabilities
\begin{equation}
  \label{eq:no-mpi}
  \prod_{i=0}^{m-1}\noem{i}\sup{MPI}(\ord_i,\ord_{i+1}), 
\end{equation}
which are easily incorporated in the trial showers described above.
We also need to include the actual MPI's. Here the philosophy
is that as soon as we have a MPI at some scale, we ignore corrections
from the full tree-level matrix element on softer jets from the
primary interaction, and allow them to be described by the PS
alone. Hence, when we start the shower from a given $m$-parton state
(with $m<N$) in step II.3, we choose the reconstructed $\ord_m$ as
starting scale. As described before, we veto any parton emission above
\ordms. However, if a MPI is generated above \ordms, it is accepted
and the shower is allowed to continue without any further
veto. (For the $m=N$ case, the shower including MPI is allowed without
restrictions, starting from $\ord_{N}$.) In this way we achieve the
same goal as in \cite{Lonnblad:2011xx}: If the $n\le N$
hardest jets in an event all belong to the primary interaction, they are
described by the tree-level ME, while all other jets are given by the
(interleaved) PS. Just as in \cite{Lonnblad:2011xx}, the
treatment of pure QCD jet production means that the Born-level cross
section is properly eikonalized by the no-MPI factor, by allowing
MPI's all the way from $\sqrt{s}$ in the trial shower for
$\noem{0}\sup{MPI}$.

\section{Results}
\label{sec:results}

We have implemented UMEPS merging in \pytppp, and will make the necessary code
public in the next major release version. 
In this section, we will concentrate on predictions for $\W$-boson and QCD 
jet production at the LHC. However, the code aims to achieve the same 
generality as the implementation of CKKW-L in \pytppp.

All input matrix element configurations are taken from Les Houches Event Files
generated with MadGraph/MadEvent, with the following settings:
\begin{itemize}
\item Fixed renormalisation scale $\mur=\mz^2$, fixed factorisation scale 
      $\muf=\mw^2$ for $\W$-production. For $2\to2$ processes in pure
      QCD, we use $\mu_{r,2\to2} = m_{\perp,1}m_{\perp,1}$ and 
      $\mu_{f,2\to2} = \min\{m^2_{\perp,1},m^2_{\perp,1}\}$.
\item CTEQ6L1 parton distributions and $\as(\mz^2) = 0.130$.
\item The merging scale $\ordms$ is defined by the minimal \pytppp evolution
      $p_{\perp,ijk}$ of all possible combinations of three partons in the
      event. $p_{\perp,ijk}$ for a single combination of three particles 
      $i$, $j$ and $k$ is defined as
      \begin{eqnarray}
        p^2_{\perp,ijk} =
        \begin{cases}
          z_{ijk}(1-z_{ijk})Q_{ij}^2 &
             ~ \textnormal{with} ~
             Q_{ij}^2 = (p_i + p_j)^2 
             ~ \textnormal{,} ~
             z_{ijk} = \frac{x_{i,jk}}{x_{i,jk}+x_{j,ik}}
             ~ \textnormal{,} ~\\
          &  x_{i,jk} = \frac{2 p_i(p_i+p_j+p_k)}{(p_i+p_j+p_k)^2}
          \quad\textnormal{for FSR}\\
          (1-z_{ijk})Q_{ij}^2 &
             ~ \textnormal{with} ~
             Q_{ij}^2 = -(p_i - p_j)^2 
             ~ \textnormal{,} ~
             z_{ijk} = \frac{(p_i - p_j + p_k)^2}{(p_i+p_k)^2}\\
          & \textnormal{for ISR}
        \end{cases}
      \end{eqnarray}
\item In QCD $2\to2$ scatterings, the kinematical transverse momentum of jets 
      is required to be larger than $p_{T,j} = 5$ GeV.
\end{itemize}
The value of $\as(\mz^2)$ was set to
match the $\as$-value obtained in fitting the PDFs used in the ME
calculation. To generate results, we have chosen the merging scale
definition to closely match the parton shower evolution
variable. The algorithm does however not depend on this particular choice.
All jets needed for analysis purposes were defined with help of 
\texttt{fastjet}-routines \cite{Cacciari:2011ma}. The momentum of the
intermediate $\W$-boson will, if required, be extracted directly from the
Monte Carlo event. We will compare UMEPS to the CKKW-L implementation in 
\pytppp. The problems we choose to highlight should be regarded as 
criticism of the implementation in \pytppp, rather than an assessment of 
CKKW-inspired methods in general.

\subsection{W-boson production}
\label{sec:results-w}

\FIGURE{
\centering
  \includegraphics[width=0.49\textwidth]{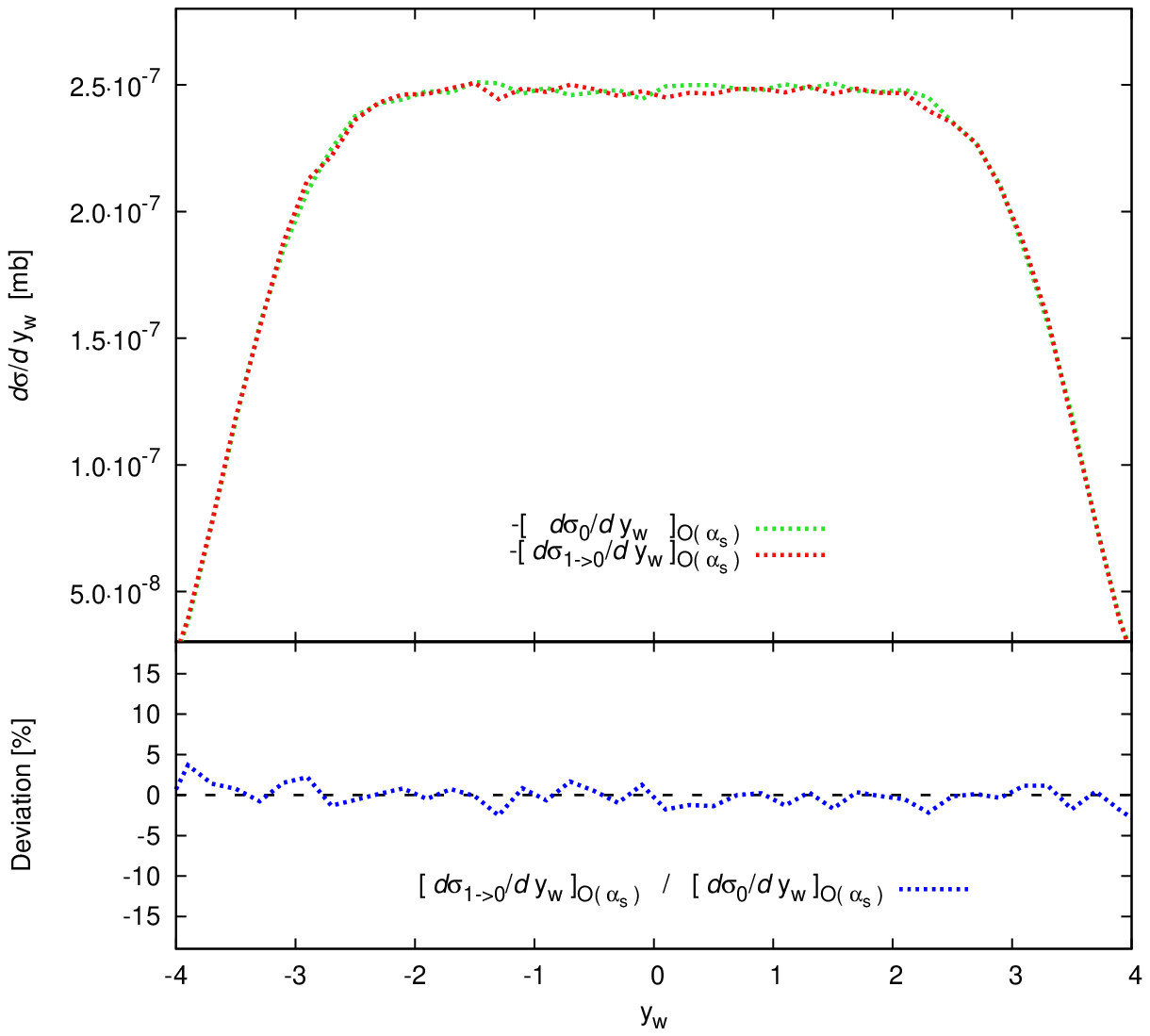}
  \includegraphics[width=0.49\textwidth]{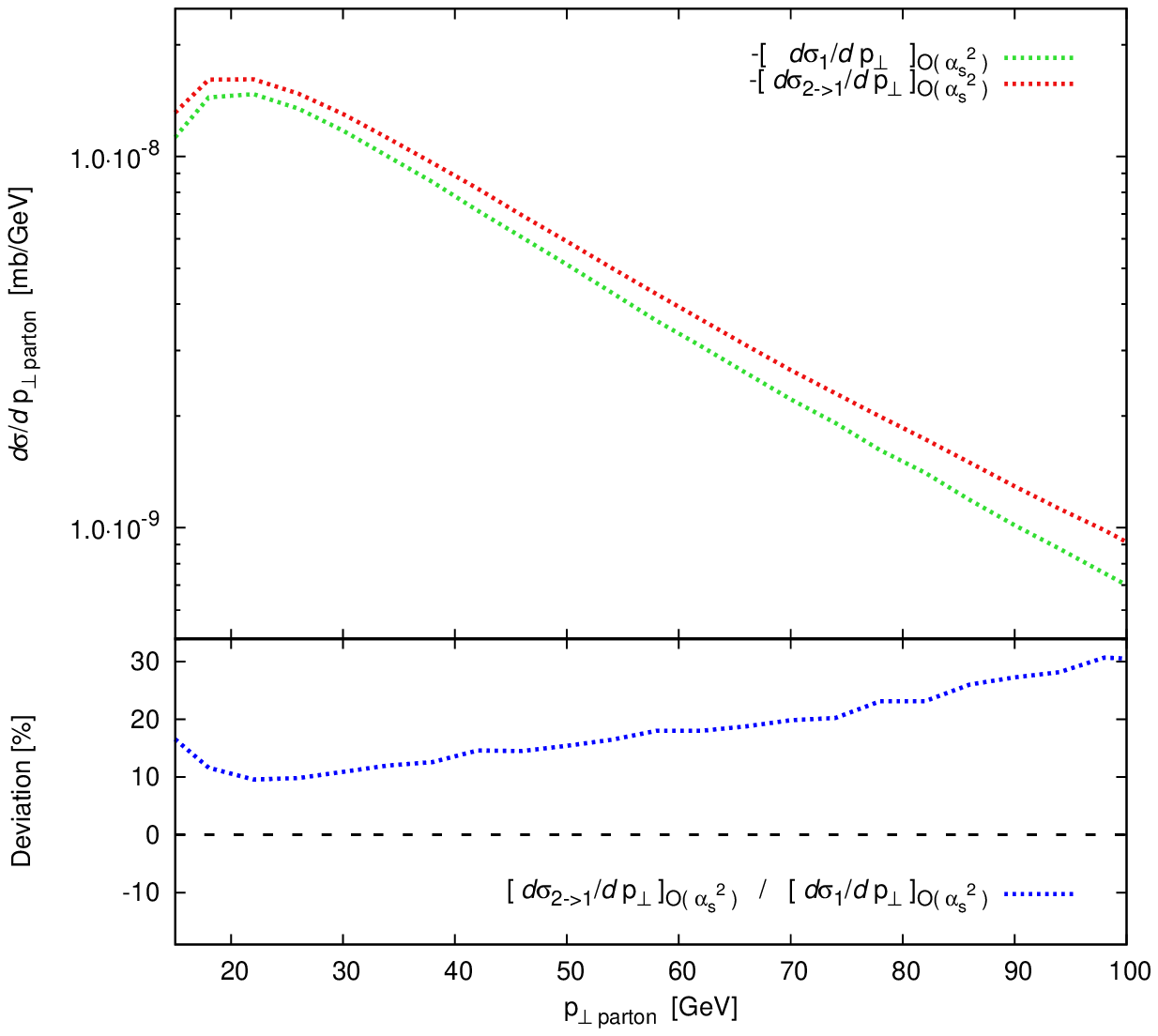}
\caption{\label{fig:w-oas-comparison}Comparison between $\Oas{}$-terms of the
  parton shower with integrated matrix elements, for $\W$-boson production in 
  $\p\p$ collisions at $\ECM=7000$ GeV. The merging scale value is 
  $\ordms = 15$ GeV. Left panel: 
  Rapidity of the $\W$-boson, intended for comparison between the integrated 
  one-jet matrix element (labelled $\left[ d\sigma_{1\rightarrow0}/dy_w \right]_{\Oas{1}}$~\!) and the 
  $\Oas{1}$-term of the no-emission probability for having no emission above
  $\ordms$ radiated off zero-jet states above
  $\ordms$ (labelled $\left[ d\sigma_{0}/dy_w \right]_{\Oas{1}}$~\!). 
  Right panel: Transverse momentum of the parton, for $\W+j$ production in 
  $\p\p$ collisions at $\ECM=7000$ GeV, intended for comparison between the 
  integrated two-jet matrix element (labelled $\left[ d\sigma_{2\rightarrow1}/d 
  p_\perp \right]_{\Oas{2}}$~\!) and the $\Oas{1}$-term of the no-emission 
  probability for having no emission above $\ordms$ radiated off one-jet states, 
  multiplying the $\W+j$ matrix element (labelled $\left[ d\sigma_{1}/d p_\perp \right]_{\Oas{2}}$~\!). }
}

We begin by comparing the result of the removal of a jet by integration with
the corresponding parton shower contribution. This is useful to assess if
performing the integration by the replacement 
$\state{n+1} \rightarrow \state{n}$ produces the desired results.

In the left
panel of Figure \ref{fig:w-oas-comparison}, we compare the integrated 
one-jet matrix element (i.e.\ the $\Oasof{1}{\mur}$-term of \eqref{eq:umeps-2j-rec1})
with the shower approximation of the $\Oasof{1}{\mur}$-term in zero-jet events. The
second term is of course just the $\Oasof{1}{\mur}$-contribution in 
\eqref{eq:me02ps-not-expanded}. The rapidity of the $\W$-boson is identical in
these two samples because \pytppp is already
matrix-element corrected for $\W+j$-states.
This demonstrates that in 
$\W$-boson production, generating the no-emission probability in zero-jet
states with \pytppp, or by a reweighted, integrated one-jet matrix element are
both legitimate ways to produce the same factor.

The right panel of Figure \ref{fig:w-oas-comparison} investigates the 
difference between the parton shower approximation of no resolved emissions
in one-jet states between the scales $\ord_1$ and $\ordms$ and the result
of constructing an unresolved emission by integrating over one parton in a
two-jet matrix element. This means that we compare the one-jet matrix element,
 multiplied by the $\Oasof{1}{\mur}$-term of the no-emission probability 
$\noem{1}(x_1,\ord_1,\ordms)$ in \eqref{eq:me12ps-not-expanded}, with the 
$\Oasof{2}{\mur}$-contribution in \eqref{eq:umeps-2j-rec2}. The comparison shows
that, as expected, the parton shower underestimates the hardness of the 
unresolved (second) emission, which is reminiscent of the fact that the 
inclusion of two-jet matrix elements into the PS prediction does in general
increase the tail of the $p_\perp$ of the hardest jet. 

\FIGURE{
\centering
  \includegraphics[width=0.49\textwidth]{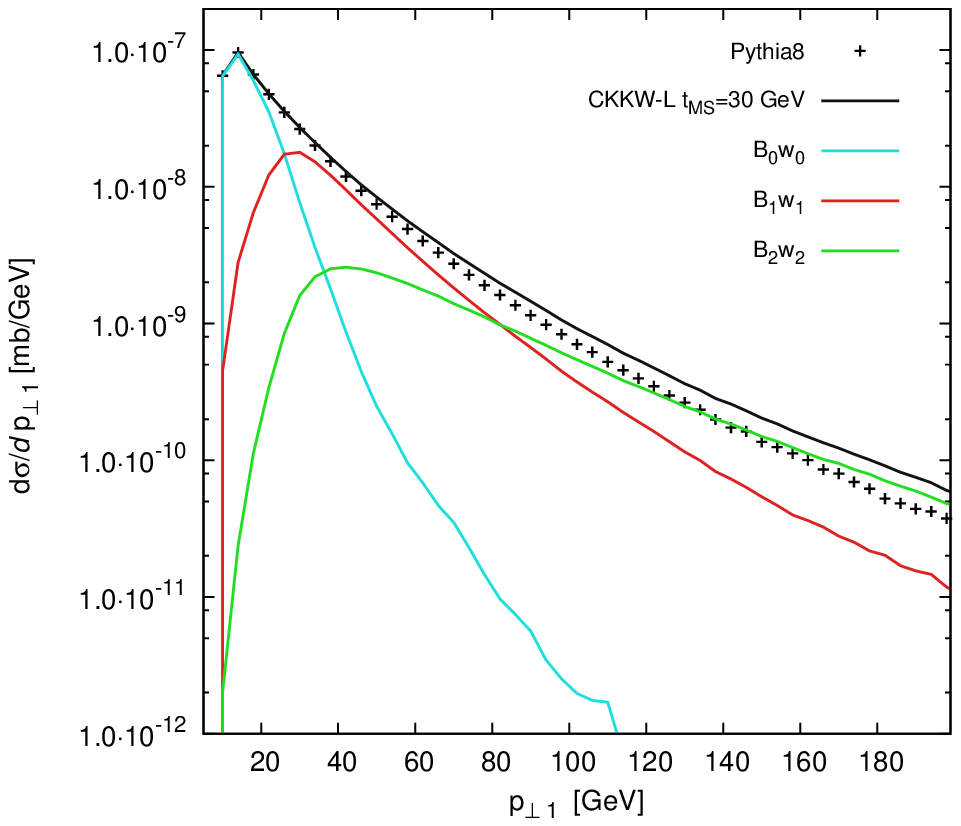}
  \includegraphics[width=0.49\textwidth]{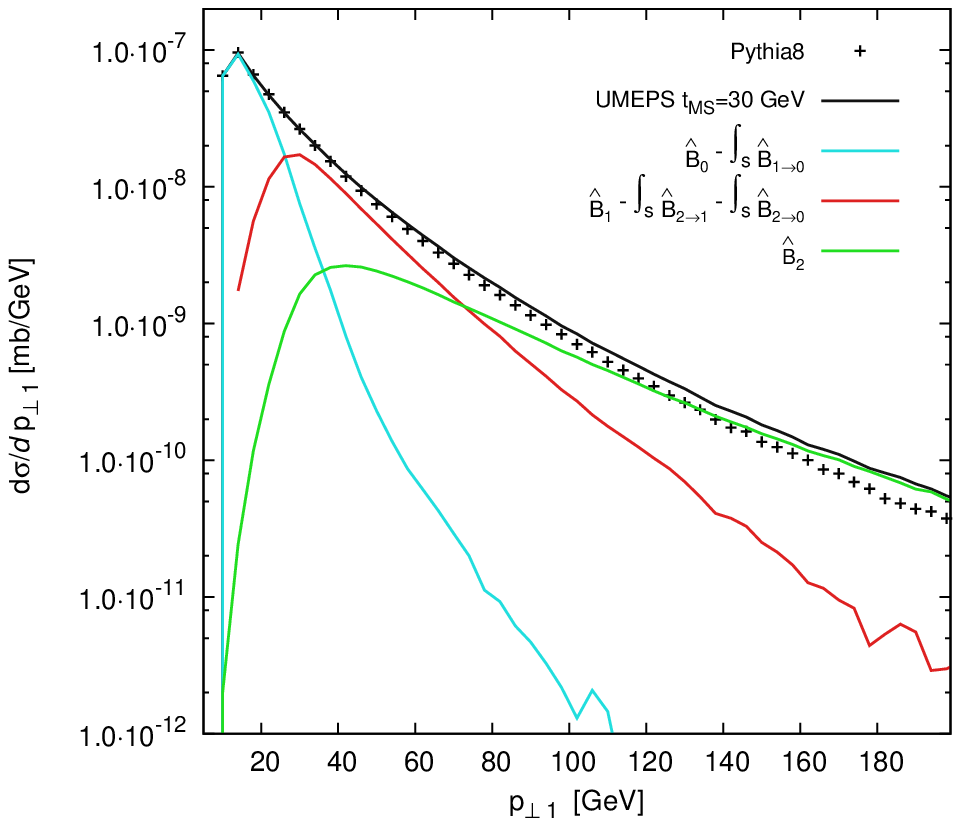}
\caption{\label{fig:w-pT1-split}Transverse momentum of the hardest jet, for $\W$-boson production in $\p\p$ collisions at $\ECM=7000$ 
  GeV, when
  merging up to two additional partons. Jets were defined with the $k_\perp$-algorithm, with $k_{\perp,min} = 10$ 
  GeV. Multi-parton interactions and hadronisation were excluded.
  Left panel: Results of the CKKW-L 
  scheme. The contributions are labelled B$_0w_0$, B$_1w_1$ and 
  B$_2w_2$ for CKKW-L-reweighted zero-, one- and two-jet matrix elements,
  respectively.
  Right panel: Results of the 
  UMEPS scheme. The contributions are labelled $\Tev{0}$, $\Tev{1}$ and 
  $\Tev{2}$ for UMEPS-reweighted zero-, one- and two-jet matrix elements,
  and $\Iev{1}{0}$ and $\Iev{2}{1}$ for UMEPS-reweighted, integrated one- and 
  two-jet samples. $\Iev{2}{0}$ indicates the two-jet contribution that was 
  integrated twice because the state $\state{1}$ after the first integration
  contained an unresolved jet.}
}

In Figure, \ref{fig:w-pT1-split} we show how matrix element samples contribute
to this increase. All jet multiplicities enter, because the merging scale is 
not defined as the jet-separation of the $k_\perp$-algorithm, and since
the merging scale cut acts on the matrix element state, while the jets are 
constructed from outgoing particles after the parton shower cascade. In CKKW-L,
the high-$p_\perp$ tail is dominated by the two-jet matrix element, with a 
major contribution from the one-jet states. The latter is significantly
lower in UMEPS, a fact that we think crucial. UMEPS correctly cancels the 
inclusion of phase space points with two resolved jets by using the two-jet
matrix element to construct a better approximation of radiating an \emph{unresolved}
parton from one-jet states. We see both in CKKW-L (left panel of Figure
\ref{fig:w-pT1-split}), and in Figure \ref{fig:w-oas-comparison}, that the 
parton shower underestimates the hardness of two-parton states. 
The description of unresolved emissions enters into the no-emission 
probabilities, with a negative $\Oas{1}$-term. Thus, the contribution 
of showered one-jet states to the tail of $p_{\perp1}$ will be larger if the 
shower description of two-jet states underestimates hardness. UMEPS improves
the description of the no-emission probability by ensuring that in inclusive
observables, resolved two-jet states are cancelled, a feature that is at work
in the tail of $p_{\perp1}$.

\FIGURE{
\centering
  \includegraphics[width=0.49\textwidth]{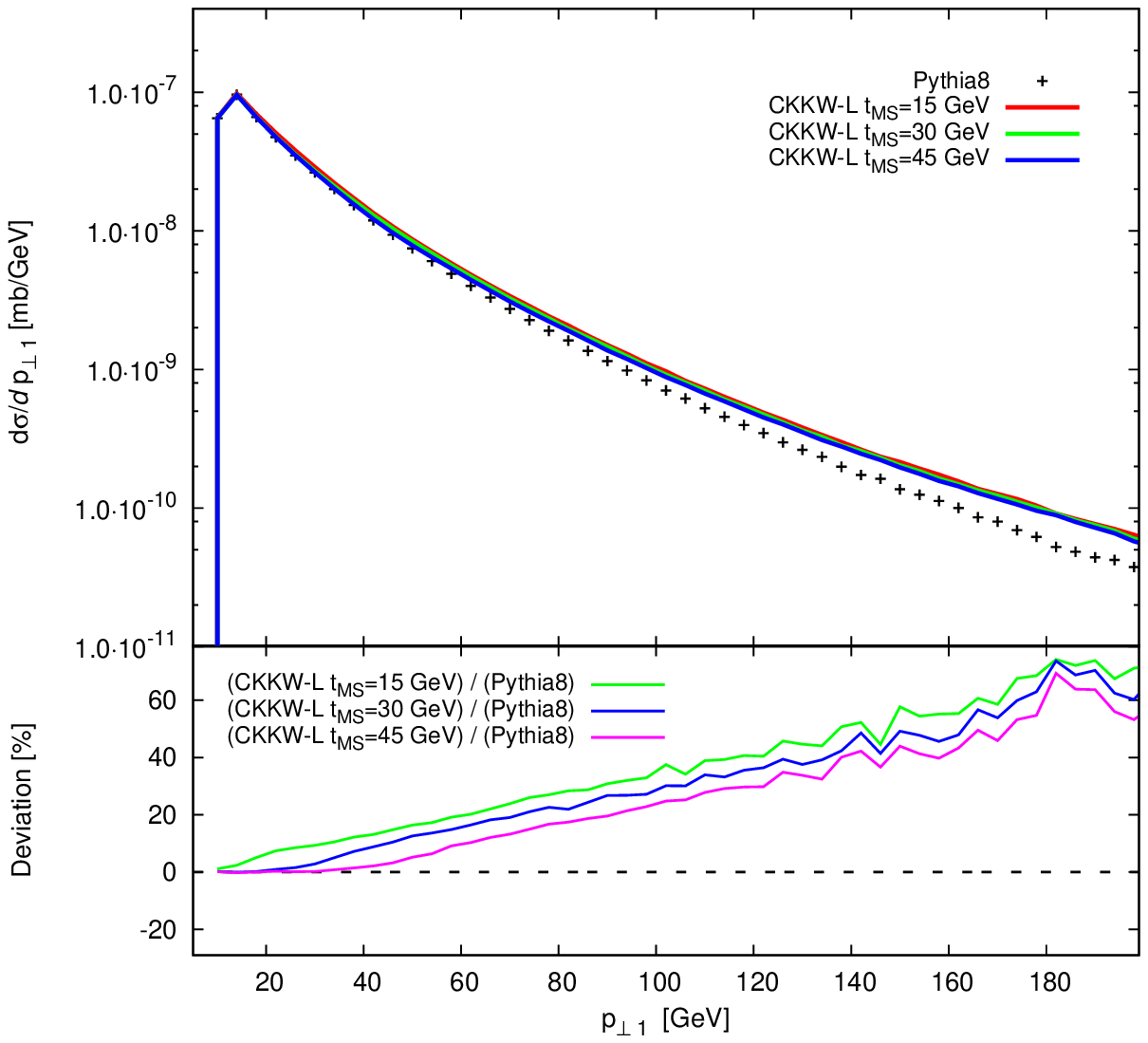}
  \includegraphics[width=0.49\textwidth]{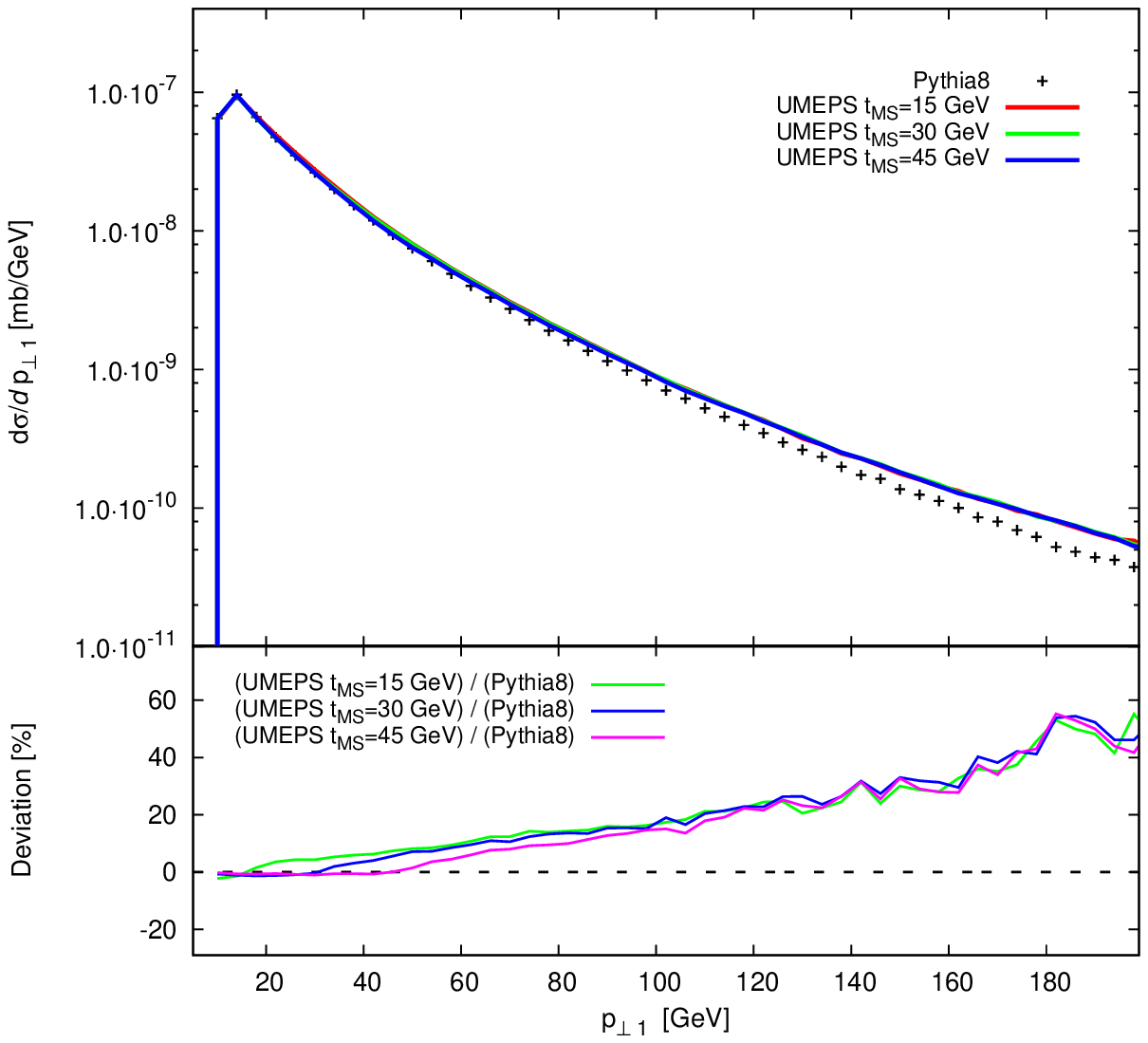}
\caption{\label{fig:w-pT1}Transverse momentum of the hardest jet, for $\W$-boson production in $\p\p$ collisions at $\ECM=7000$ 
  GeV, when
  merging up to two additional partons. Jets were defined with the $k_\perp$-algorithm, with $k_{\perp,min} = 10$ 
  GeV. Multi-parton interactions and hadronisation were excluded. The lower insets show the deviation of merged results from default
  \pytppp, for three different $\ordms$-values. Left panel: Results of
  the CKKW-L scheme. Right panel: Results
  of the UMEPS scheme. }
}

Variations in the description of $p_{\perp1}$ are also visible in
Figure \ref{fig:w-pT1}, where we show the transverse momentum of the
hardest jet in CKKW-L and UMEPS\footnote{Note that the co-variation of merged 
results in the ratio inset is due to fluctuations in the \pytppp reference 
curve.}. The trend sketched in the previous
paragraph is particularly clear in the insets comparing to default
\pytppp: UMEPS produces a softer tail in $p_{\perp1}$ than CKKW-L. The
harder tail in CKKW-L is due to a worse description of unresolved
emissions. It is fair to say that the difference between CKKW-L and
UMEPS hints at the size of relic effects from not cancelling the
higher-multiplicity matrix elements in a well-defined way.  Merging
scale variations in tree-level merging schemes arise from a mismatch
of unresolved emissions exponentiated in no-emission probabilities and
tree-level matrix elements for hard, resolved jets. UMEPS has a
significantly lower merging scale variation since the method enforces
a cancellation of resolved and unresolved contributions.

\FIGURE{
\centering
  \includegraphics[width=0.49\textwidth]{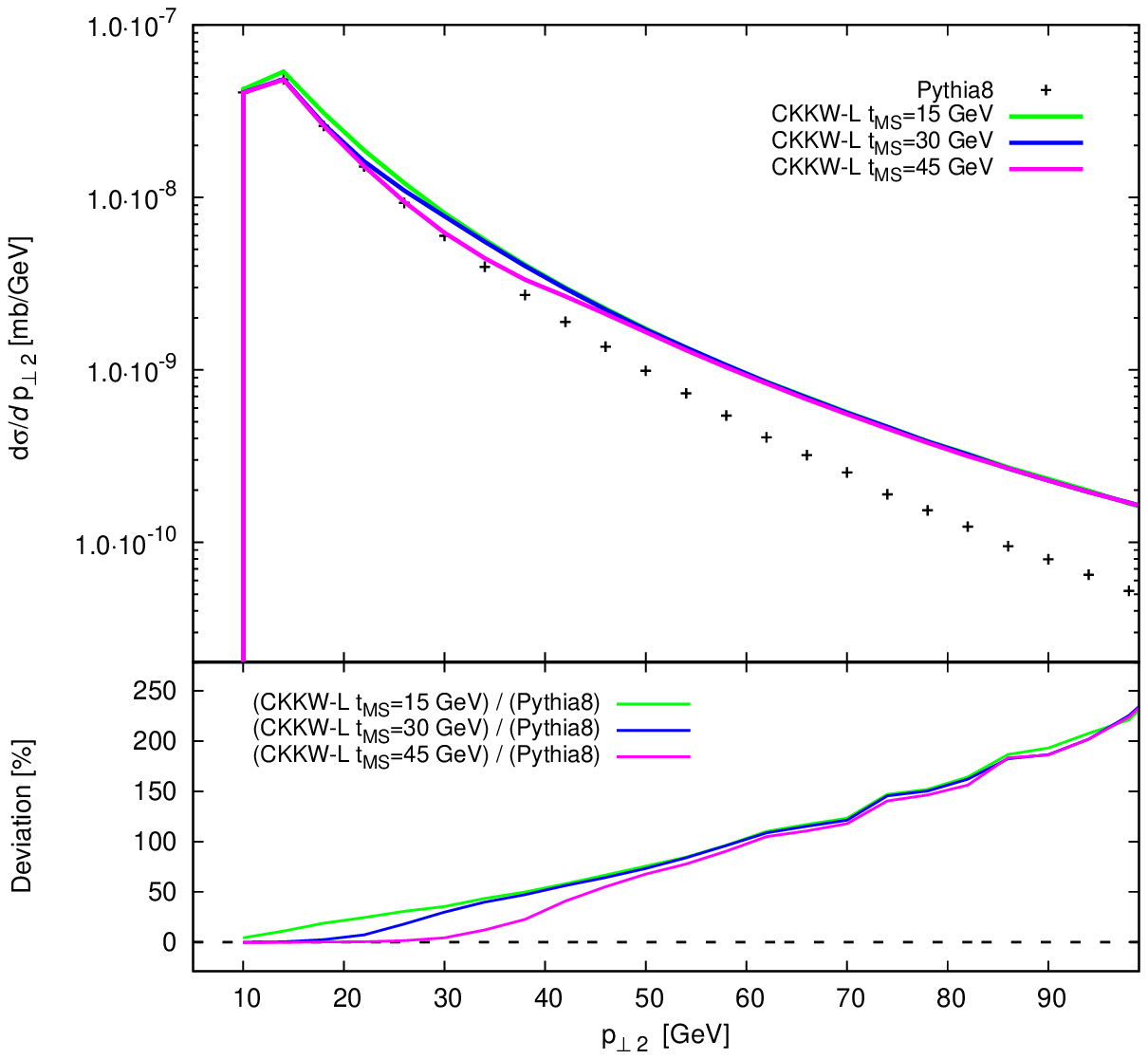}
  \includegraphics[width=0.49\textwidth]{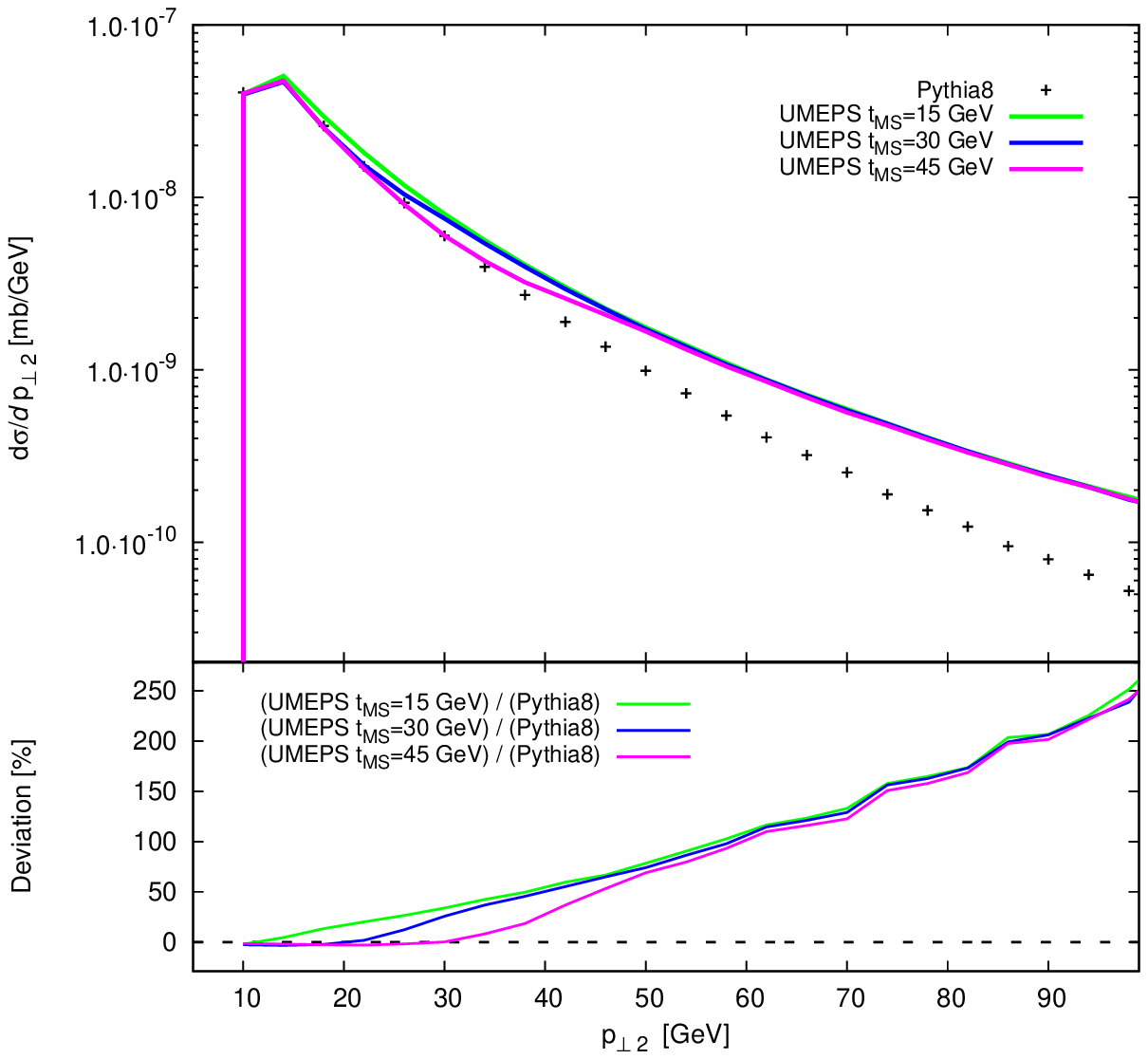}
\caption{\label{fig:w-pT2}Transverse momentum of the second-hardest jet, for $\W$-boson production in $\p\p$ collisions at 
  $\ECM=7000$ GeV. Jets were defined with the $k_\perp$-algorithm, with $k_{\perp,min} = 10$ 
  GeV, when
  merging up to two additional partons. Multi-parton interactions and hadronisation were excluded. The lower insets show the deviation of merged results 
  from default \pytppp, for three different $\ordms$-values. Left panel: 
  Results of the CKKW-L scheme. Right 
  panel: Results of the UMEPS scheme.}
}

Figure \ref{fig:w-pT2} shows that for very exclusive observables, CKKW-L and
UMEPS are virtually indistinguishable. In this example, this is of course 
expected since the treatment of the highest multiplicity (here, the two-jet)
matrix element is identical for both cases.

\FIGURE{
\centering
  \includegraphics[width=0.6\textwidth]{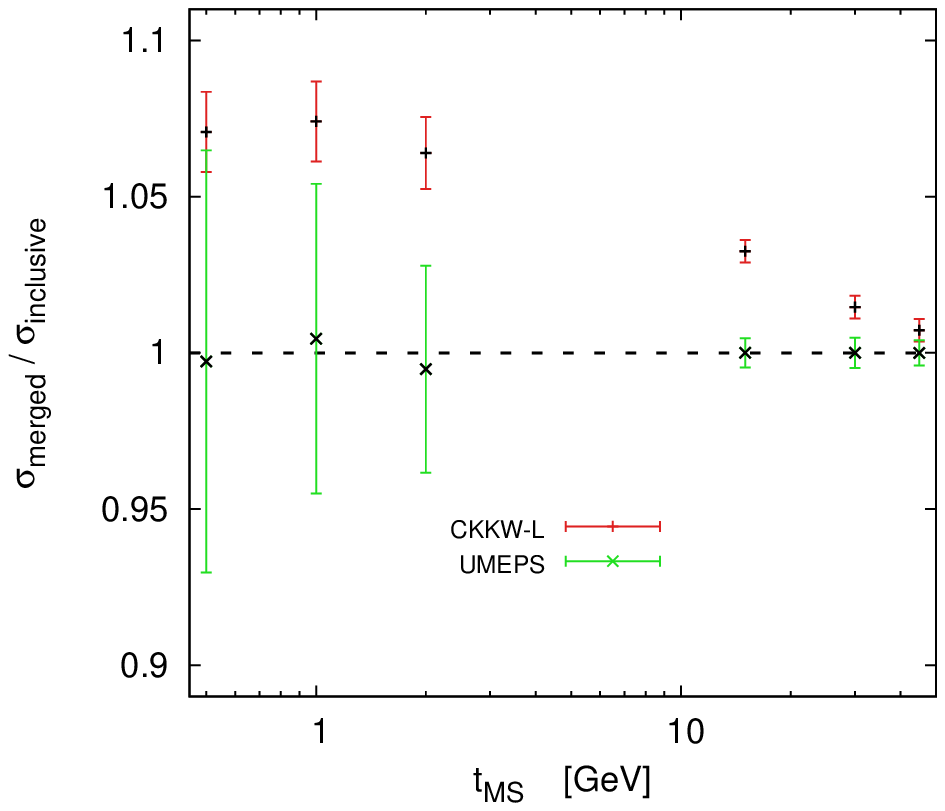}
  \caption{\label{fig:xsections} Inclusive cross section for UMEPS-
    and CKKW-L merging of up to two additional jets in $\W$-boson
    production in $\p\p$ collisions at $\ECM=7000$ GeV (labelled
    $\sigma_{\textnormal{merged}}$), in comparison to the the
    lowest-multiplicity inclusive cross section
    $\sigma_{\textnormal{inclusive}}$. The error bars represent only
    the statistical error on the merged cross section. For the UMEPS
    sample this becomes large for small merging scales, but as it is
    the same samples which are added and the subtracted, the central
    value stays very close to unity.}
}

We would now like to perform a stress-test of the merging scale dependence.
Since UMEPS properly cancels the effects of adding multi-jet matrix elements
by subtracting their integrated counter-parts, it is in principle possible to
push the merging scale to very small values. The variation of the 
inclusive cross section is shown in Figure \ref{fig:xsections}. It is clear 
that UMEPS does indeed preserve the inclusive cross section, while for 
CKKW-L, very small merging scales lead to large changes, rendering the 
method unreliable. However, the error convergence in UMEPS is, due to the 
negative weights, significantly slower. We will comment on this below.

\FIGURE{
\centering
  \includegraphics[width=0.49\textwidth]{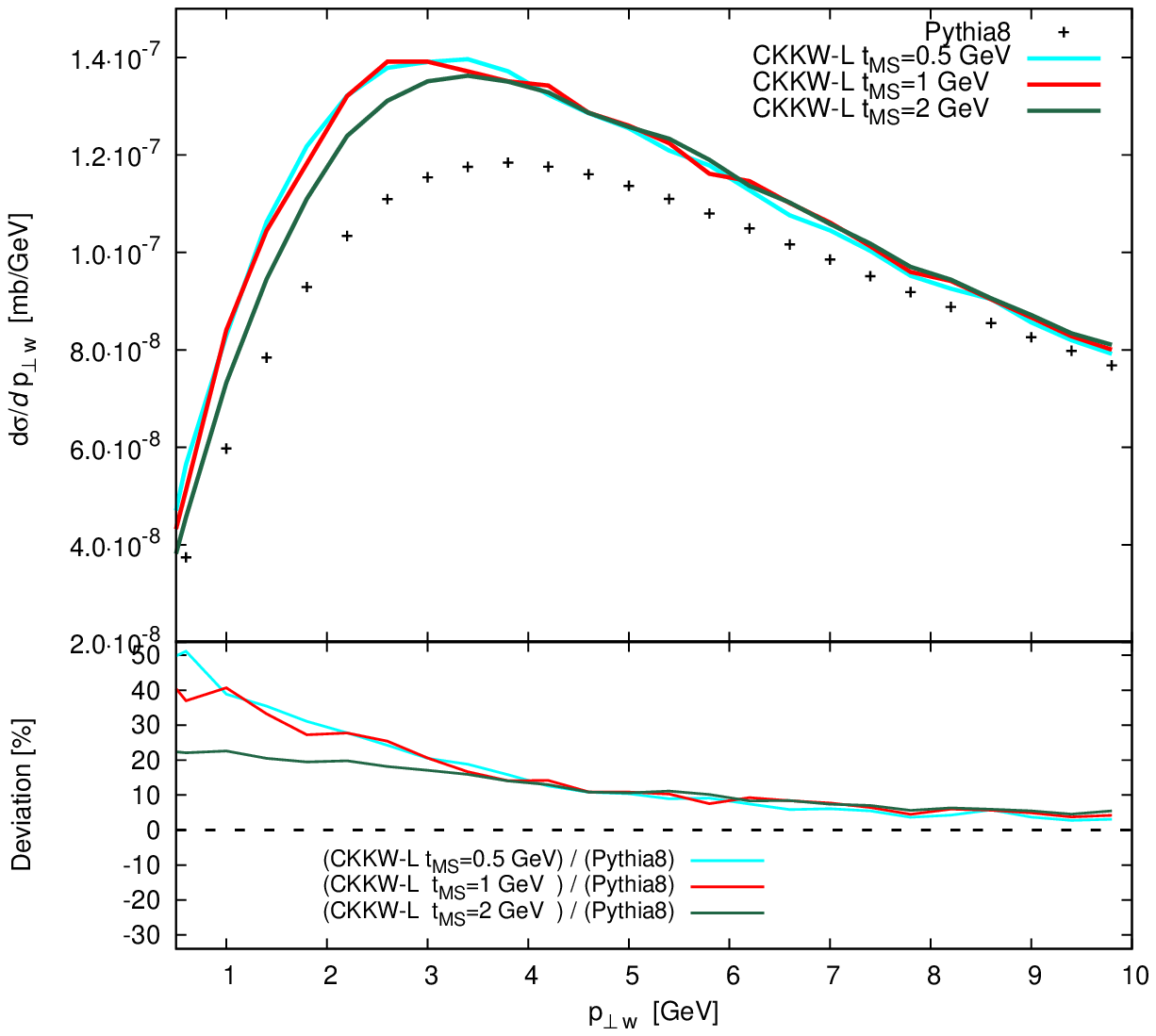}
  \includegraphics[width=0.49\textwidth]{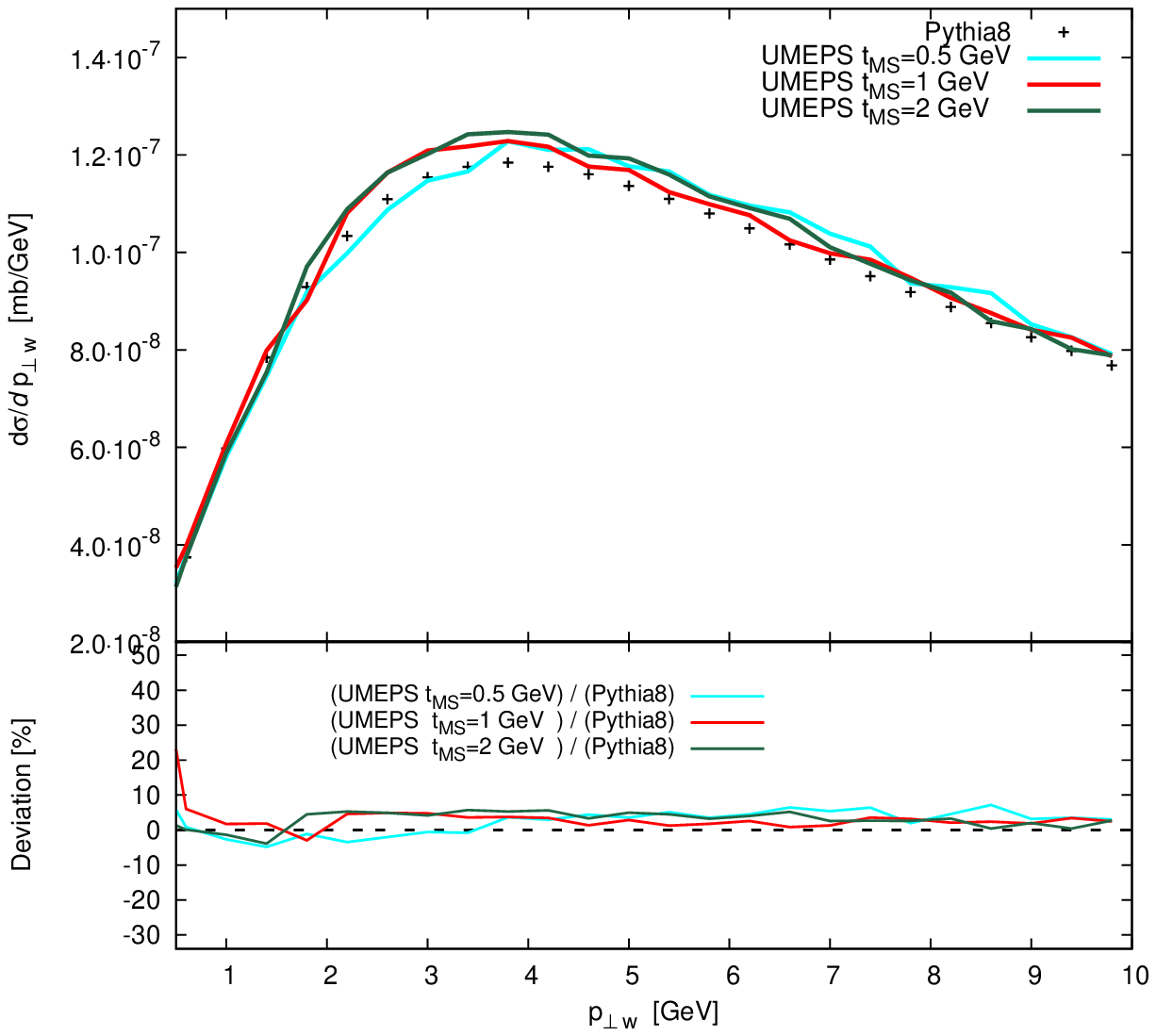}
\caption{\label{fig:pTw-low-scale}Transverse momentum of the $\W$-boson, for
  $\W$-boson production in $\p\p$ collisions at $\ECM=7000$ GeV, when
  merging up to two additional partons. Multi-parton interactions and hadronisation were excluded. The \pytppp 
  results are generated with the default
  settings, in particular with a primordial transverse momentum of 
  $k_{\perp, p} = 2$ GeV. All merged curves have been generated with 
  $k_{\perp, p} = 0.5$ GeV.
  The lower insets show the deviation of merged results 
  from default \pytppp, for three different $\ordms$-values. Left panel: 
  Results of the CKKW-L scheme. Right 
  panel: Results of the UMEPS scheme.}
}

These unitarity violations might not induce drastic effects in the
description of hard-scale observables like the transverse-momentum
distribution of the $\W$-boson.
However, magnifying the low-scale
description of this observable (Figure \ref{fig:pTw-low-scale})
reveals problems. Figure \ref{fig:pTw-low-scale} serves two
purposes. It clearly shows that by pushing the merging scale to small
values in CKKW-L, sub-leading contributions in the multi-jet matrix
elements start to contribute more. Since those sub-leading
contributions cannot be cancelled by the default parton shower, major
increases over \pytppp are found. UMEPS explicitly cancels these
sub-leading terms, and thus leads to a reliable prediction.  The
second observation in Figure \ref{fig:pTw-low-scale} is the dependence
on the primordial transverse momentum parameter $k_{\perp,p}$. This
parameter was introduced in event generators to account for the
transverse momentum of partons in the incoming protons, which cannot
be generated in initial state DGLAP evolution. If this were the only
effect to be modelled by $k_{\perp,p}$, a value of $k_{\perp,p}\approx
0.3$ GeV per incoming parton would seem appropriate. However, in
current event generator tunes, significantly higher values
($k_{\perp,p}\approx 2$ GeV) are required
\cite{Miu:1998ju,Balazs:2000sz}, potentially to compensate for an
incomplete phase space coverage in initial state showers due to the
shower cut-off. The value of $k_{\perp,p}$ is mainly fixed by tuning
to the position of the peak of the transverse momentum spectrum of the
$\Z$- or $\W$ boson. Increasing the value of $k_{\perp,p}$ roughly
corresponds to pushing the peak to higher $p_\perp$ values. Figure
\ref{fig:pTw-low-scale} compares the UMEPS and CKKW-L predictions for
the transverse momentum of the $\W$ boson, with $k_{\perp,p}=0.5$ GeV,
to default, tuned \pytppp with $k_{\perp,p}=2.0$ GeV. Unitarity
violations in CKKW-L pull the peak back to lower $p_\perp$. This fact
is virtually unchanged if we had used $k_{\perp,p}=2.0$ for CKKW-L
predictions instead, suggesting that if we positively wanted to use a
very low merging scale, an increase in $k_{\perp,p}$ would be
necessary. UMEPS on the other hand can be used with very low merging
scales, and in particular shows the interesting feature of matching
the default \pytppp curve \emph{without} having a high $k_{\perp,p}$
value. We believe this is due to a better modelling of logarithms of
the form $\ln\left(1/x\right)$, which are present in the matrix
element, and which are included in a unitary way in UMEPS -- allowing
for a much more natural value of $k_{\perp,p}$. This result
is of course very preliminary, since there are e.g.\ correlations of
the shower cut-off $p_{\perp \min}$ and $k_{\perp,p}$. One would hope
that matrix-element merging would allow to lower $p_{\perp \min}$,
which might mean having to make a compromise for the value of
$k_{\perp,p}$. We will come back to these aspects when presenting
tunes for matrix-element-merged \pytppp in a future publication.

\subsection{Dijet production}
\label{sec:results-dijet}

We would further like to mention QCD dijet production at the LHC, 
since potential merging scale dependencies enter already when merging 
dijet- and three-jet matrix elements, and to demonstrate the flexibility of 
our implementation.
The main objective of including QCD dijet production in this publication was 
to assess the treatment of MPI discussed at the end of 
section \ref{sec:umeps-step-by-step}. This is most effectively done by 
comparing to data, and before these we would like to only stress one issue.

\FIGURE{
\centering
  \includegraphics[width=0.49\textwidth]{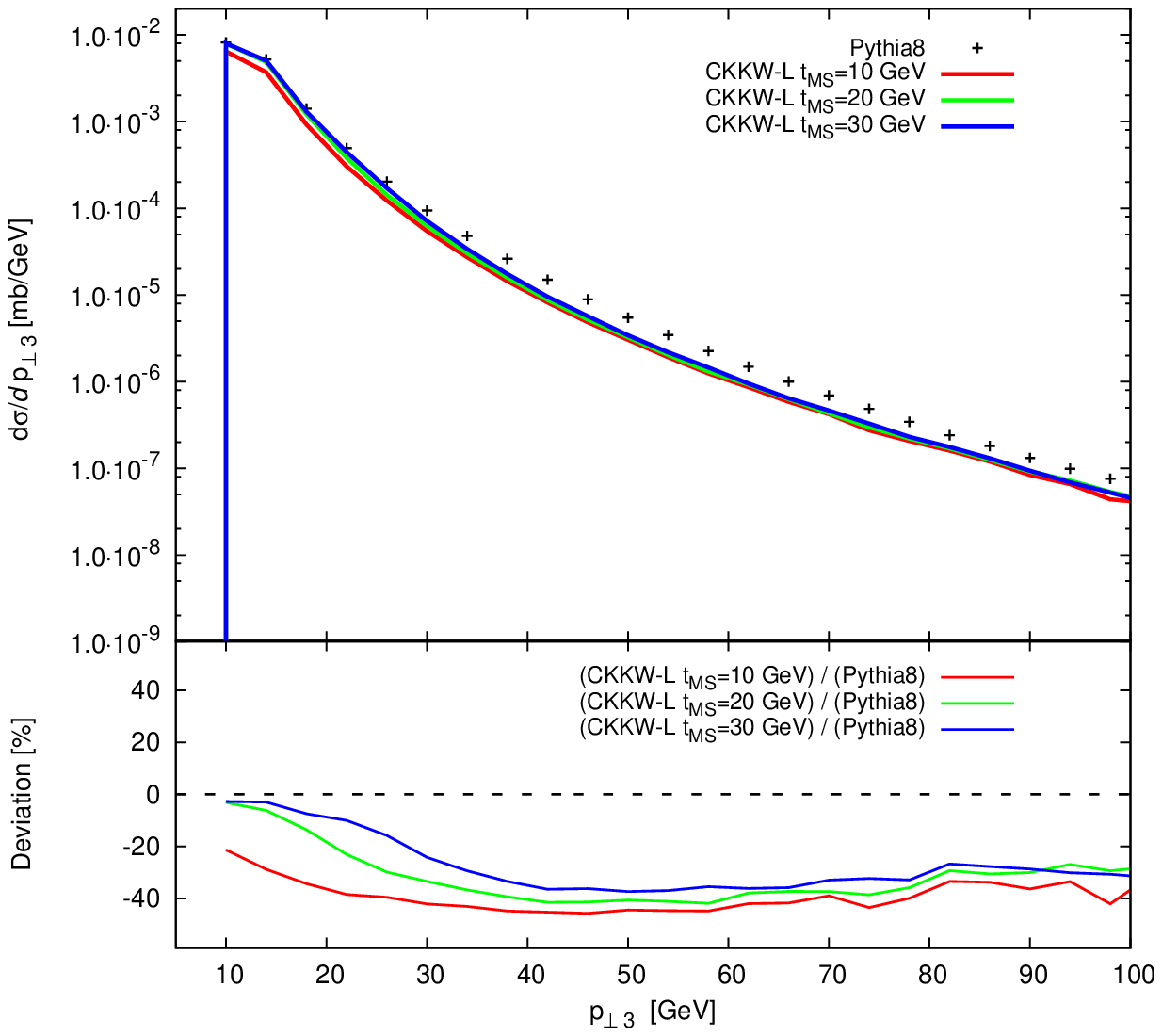}
  \includegraphics[width=0.49\textwidth]{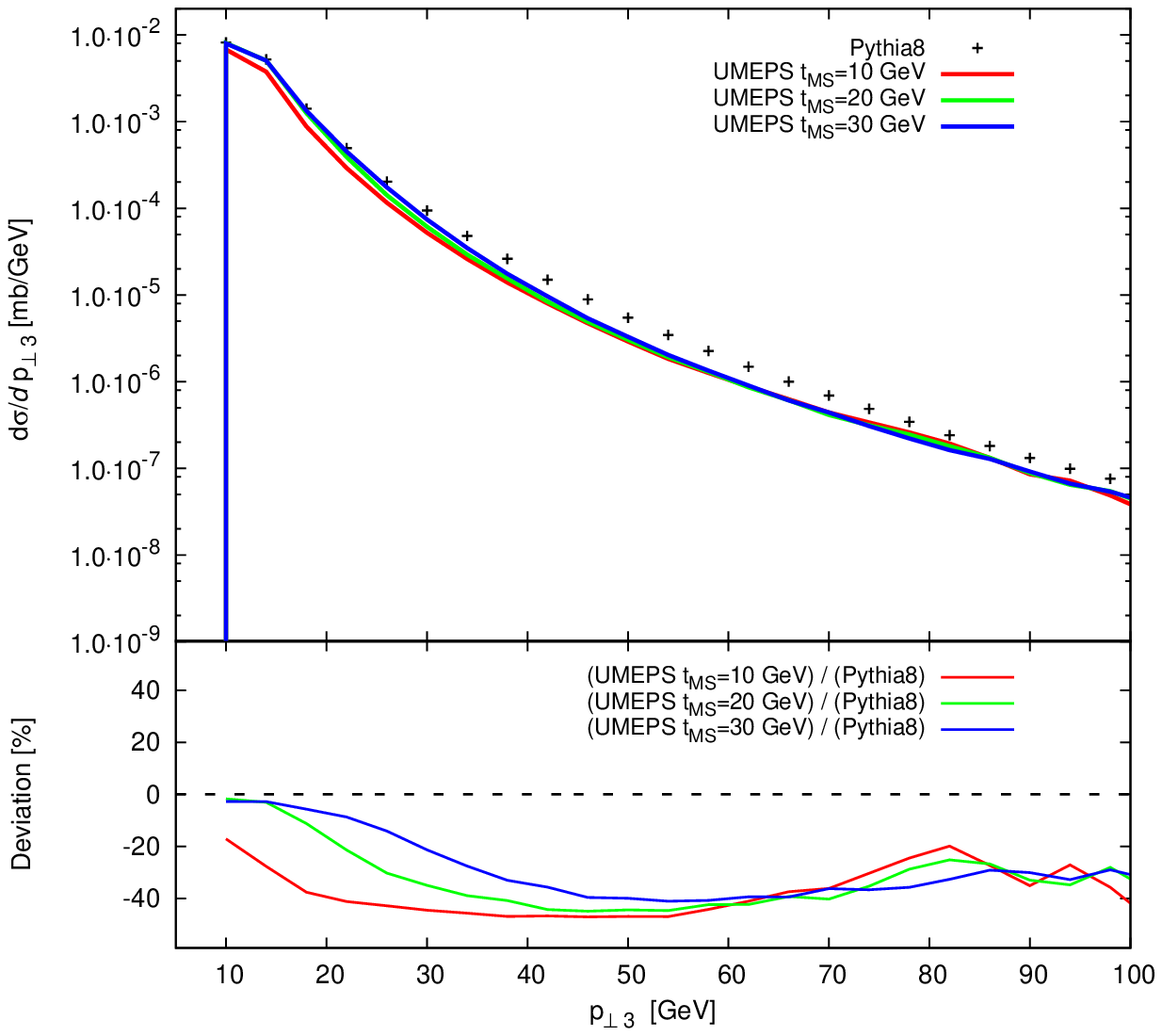}
\caption{\label{fig:qcd-pT3} Transverse momentum of the third hardest jet,
  for pure QCD dijet production in $\p\p$ collisions at $\ECM=7000$ GeV, when
  merging up to two additional partons.
  Jets were defined with the $k_\perp$-algorithm, with $k_{\perp,min} = 10$ 
  GeV. Multi-parton interactions and hadronisation were excluded. The lower insets show the deviation of merged results 
  from default \pytppp, for three different $\ordms$-values. Left panel: 
  Results of the CKKW-L scheme. Right 
  panel: Results of the UMEPS scheme.}
}

Figure \ref{fig:qcd-pT3} shows the transverse momentum of the
third-hardest jet. We see that UMEPS and CKKW-L show very similar
changes when compared to \pytppp. However, neither curves show the
high-$p_\perp$ increase seen in \cite{Lonnblad:2011xx}. This is simply
because we have revised the choice of the renormalisation scale in the
core $2\to2$ QCD scattering.  In \cite{Lonnblad:2011xx}, the two
powers of $\as(\mur)$ in the core $2\to2$ process were never touched,
and thus were evaluated with the rather unfortunate choice $\mur=\mz^2$
in the input Les Houches events. This has been rectified in the
current release of (CKKW-L in) \pytppp, i.e.\ the scale choice
($\mu_{r,2\to2} = m_{\perp,1}m_{\perp,1}$) is
now dynamical. Potential reweighting (due to the usage of fixed $\mur$
in the LHEF generation is handled internally in \pytppp. The trend
that pure QCD multi-jet matrix elements have a softer spectrum of 
well-separated jets has already been observed 
in \cite{Lonnblad:2011xx,Corke:2010yf}. The merging scale variation in UMEPS
is within the statistical error of the samples. The statical uncertainty is 
larger in UMEPS than in CKKW-L, due to cancellations between positive and 
negative weights (see the last part on section \ref{sec:discussion}).

\subsection{Comparison with data}
\label{sec:comparison-with-data}

In this section, we would like to confront UMEPS with experimental data. 
Event generator predictions were obtained with the settings of Tune A2 \cite{ATL:2012-003pub}.
The results should of course not be regarded as final statement, since changes in the perturbative physics
in event generators in principle request a full re-tuning. The intention of
this section is to investigate if after including matrix-element information, 
hard-scale features are closer to measurements, and to assess the changes in
underlying event description. All plots were produced with \rivet 
\cite{Buckley:2010ar}. We apologise if the selection of experimental 
measurements seems biased.

\FIGURE{
\centering
  \includegraphics[width=0.49\textwidth]{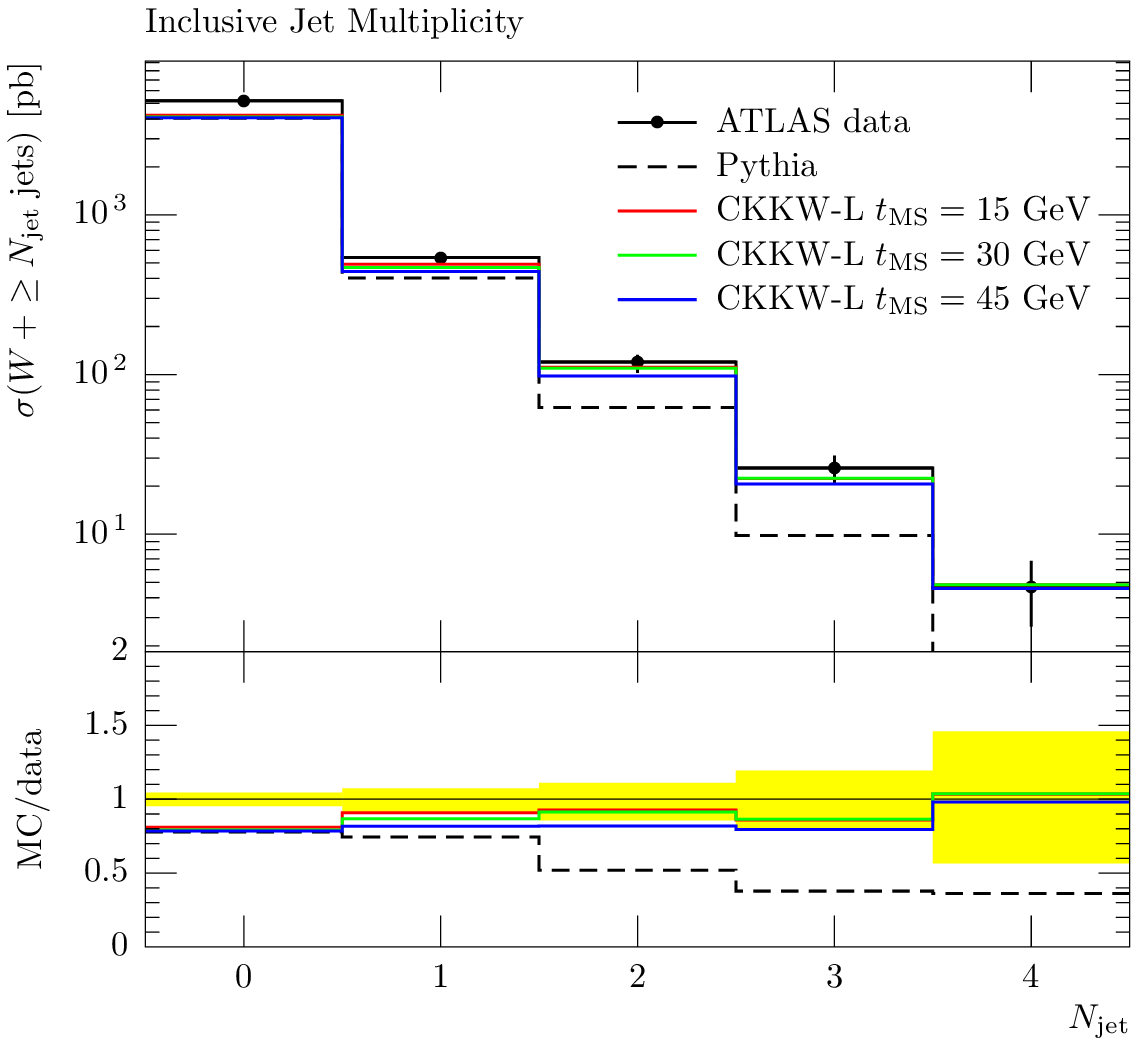}
  \includegraphics[width=0.49\textwidth]{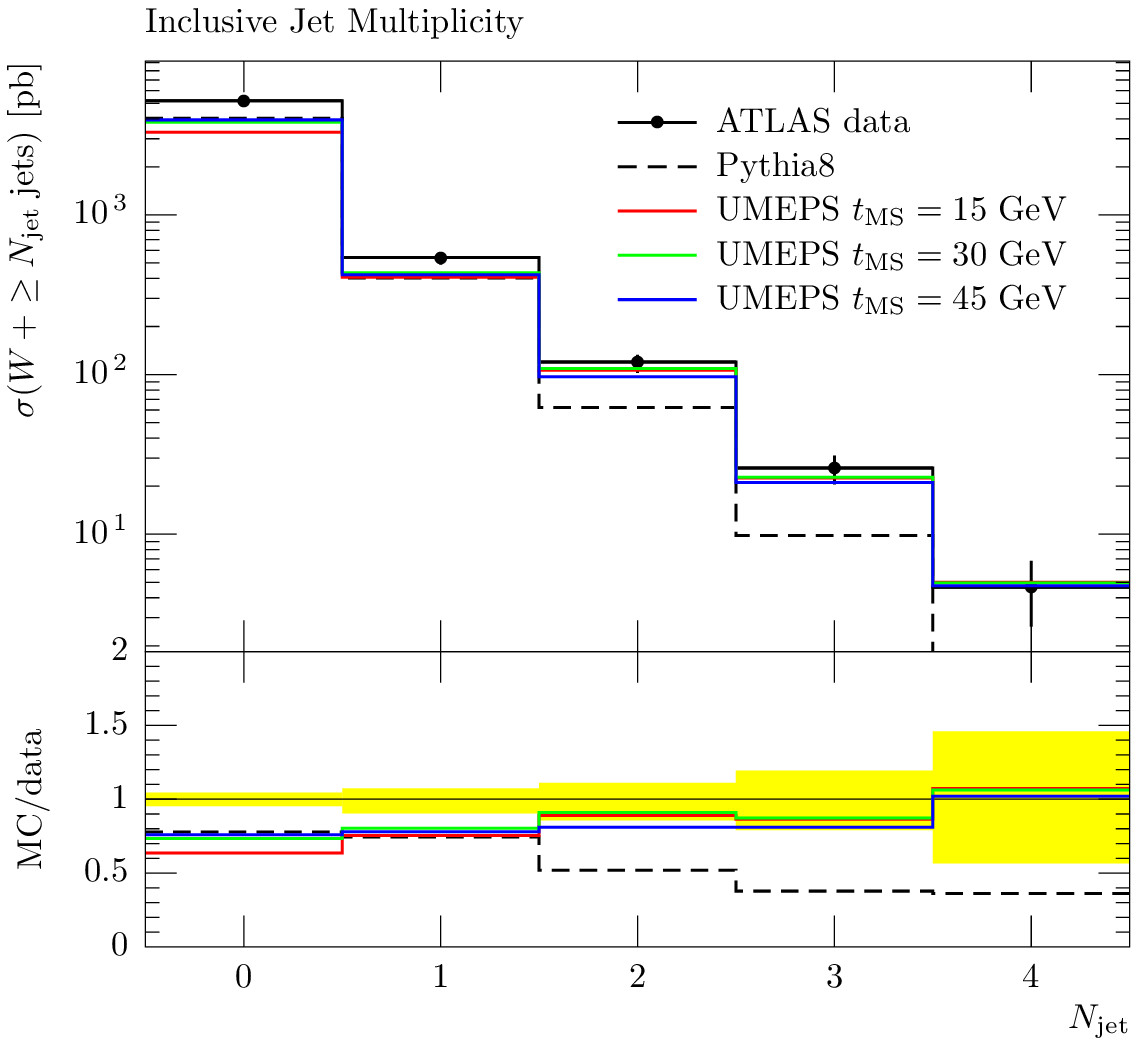}
  \caption{\label{fig:atlas-w-njet-rates}Jet multiplicity in $\W$-boson 
  production, for three different merging scales, as measured by 
  ATLAS \cite{Aad:2012en}. Effects of multiple scatterings and 
  hadronisation are included.}
}

In Figure \ref{fig:atlas-w-njet-rates}, we show jet rates in $\W$-boson production at
the ATLAS \cite{Aad:2012en}. We find an improved description after including up to two 
additional jets, and little differences between UMEPS and CKKW-L.

\FIGURE{
\centering
  \includegraphics[width=0.49\textwidth]{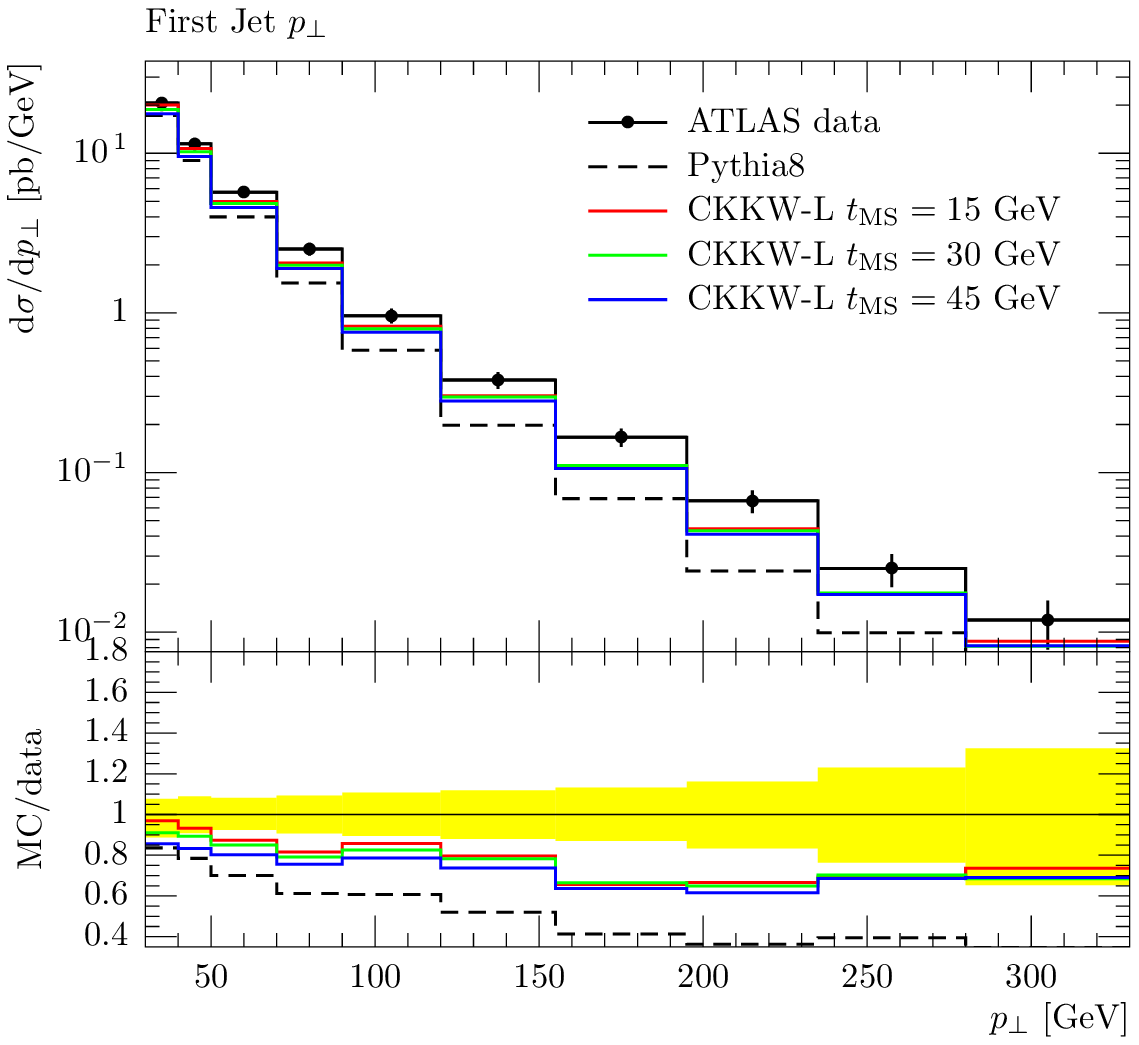}
  \includegraphics[width=0.49\textwidth]{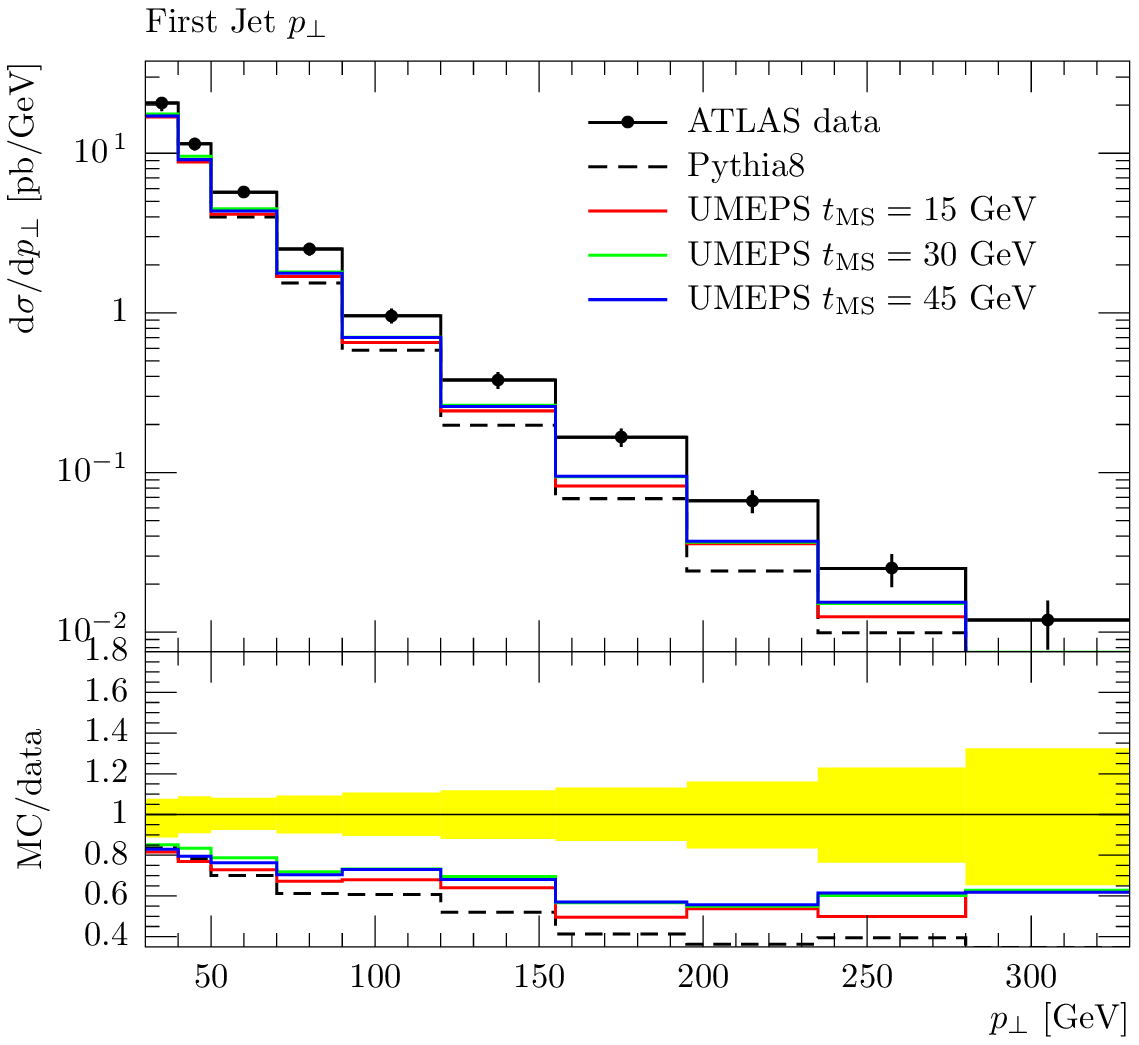}
  \caption{\label{fig:atlas-w-pt-first}Transverse momentum of the hardest jet  in $\W$-boson 
  production, for three different merging scales, as measured by 
  ATLAS \cite{Aad:2012en}. Effects of multiple scatterings and 
  hadronisation are included.}
}

The transverse momentum of the hardest jet is shown in Figure \ref{fig:atlas-w-pt-first}. Again, we find that the
shape description is improved by the CKKW-L and UMEPS methods. CKKW-L shows 
merging scale variations at lower $p_\perp$-values, since a slightly different inclusive cross 
section for a low merging scale leads to a slightly different normalisations. UMEPS
on the other hand suffers from statistical fluctuations for a low merging scale value ($\ordms=15$ GeV),
while the curves for $\ordms=30$ GeV and $\ordms=45$ completely overlap.
Note that the $p_\perp$ spectrum of UMEPS is a little softer than CKKW-L, in 
accordance with Figure \ref{fig:w-pT1}.

\FIGURE{
\centering
  \includegraphics[width=0.49\textwidth]{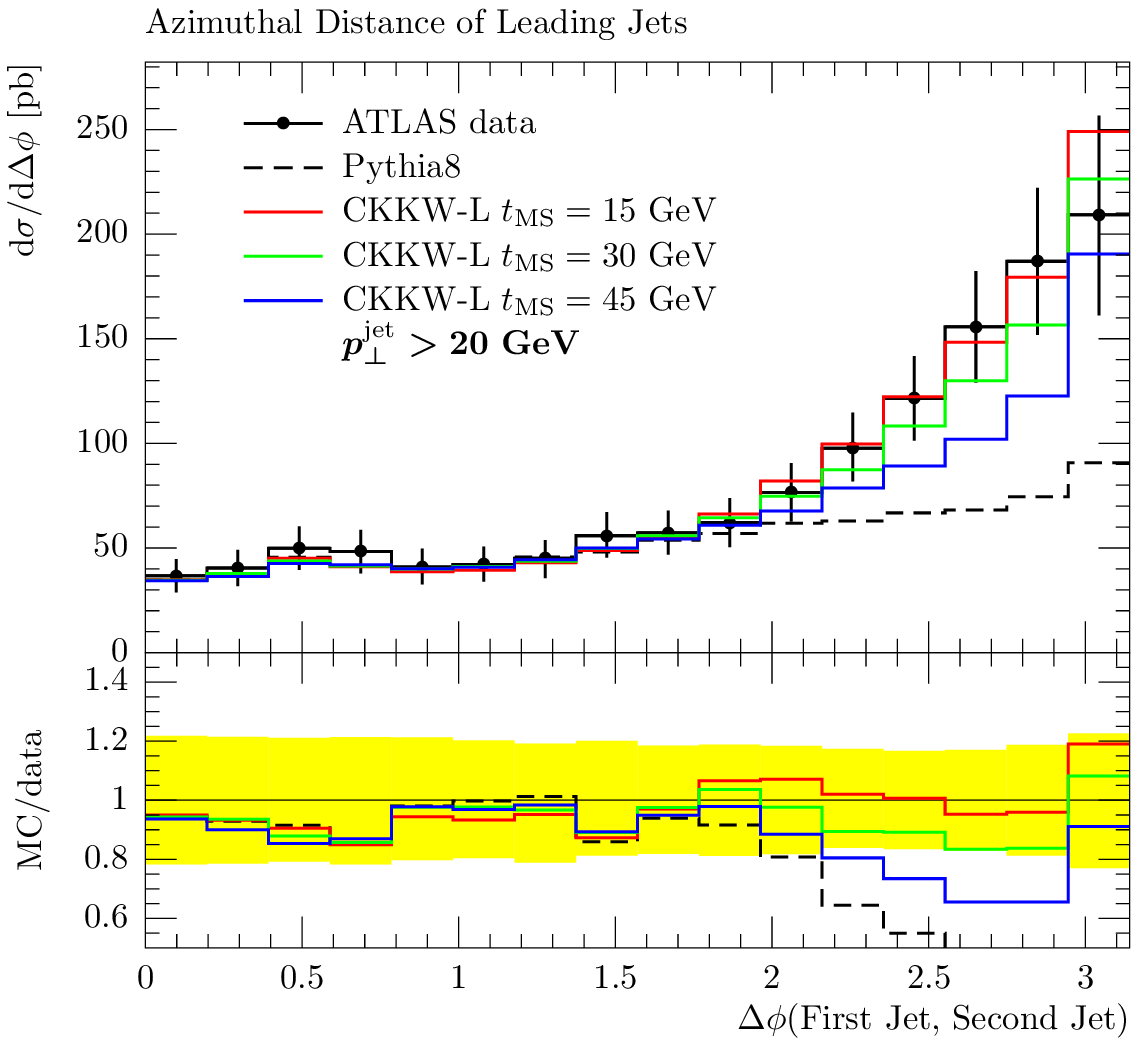}
  \includegraphics[width=0.49\textwidth]{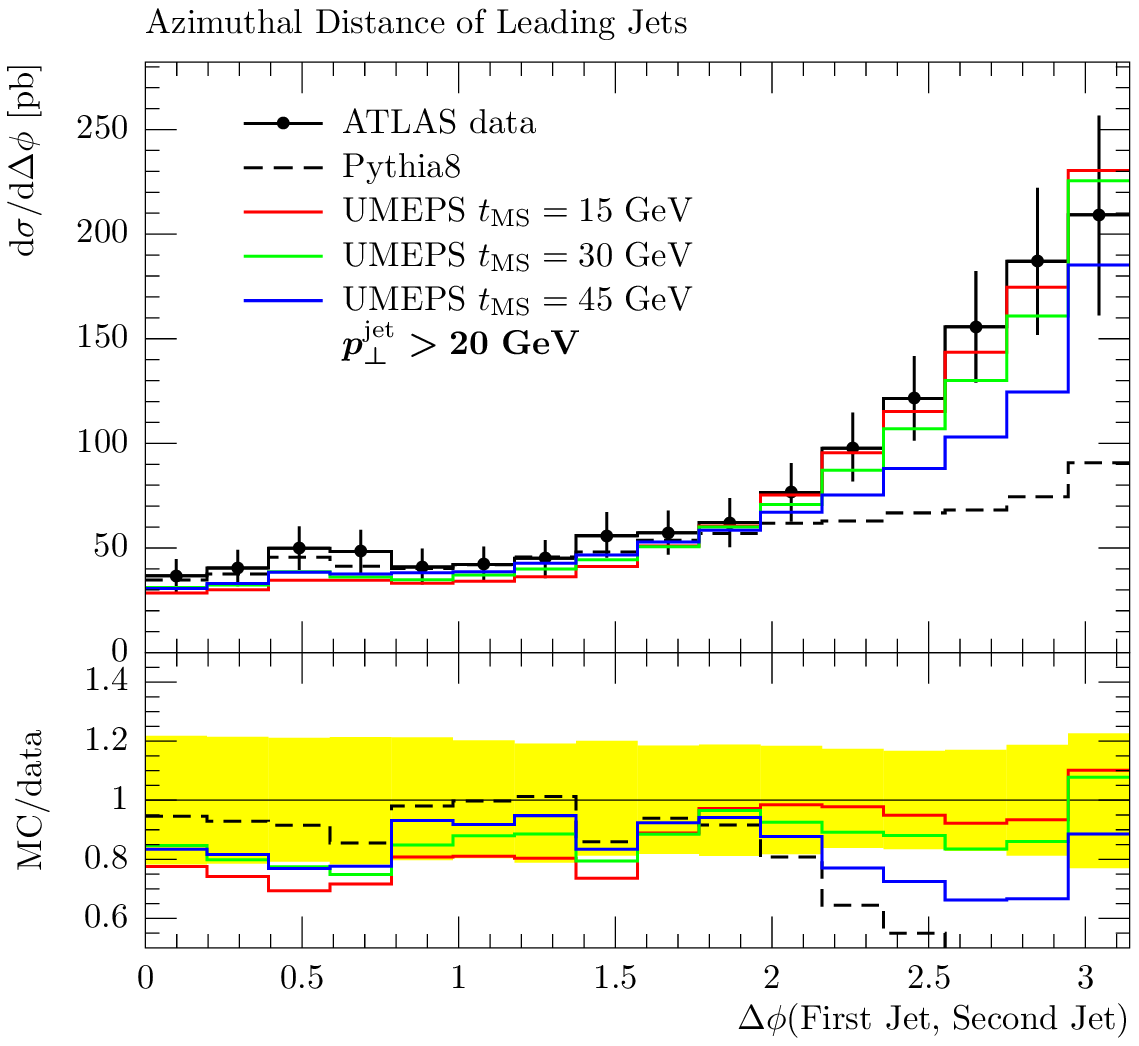}
  \caption{\label{fig:atlas-w-dphi12}Azimuthal distance between the two 
  hardest jets in $\W$-boson production, for three different merging scales, 
  as measured by ATLAS \cite{Aad:2012en}. Effects of multiple scatterings and 
  hadronisation are included.}
}

It is interesting to investigate when tree-level matrix element merging 
schemes produce large uncertainties. Figure \ref{fig:atlas-w-dphi12}
shows the azimuthal distance $\Delta\phi_{12}$ between the two hardest jets.
The parton shower alone cannot describe the peak at $\pi$. If the merging 
scale is low, the two-jet matrix element will give the dominant contribution 
in the peak region. High merging scales will increase the influence of the 
shower, thus degrading the description of the peak. Thus, this observable 
carries major merging scale variations, and provides excellent guidance for 
future improvements.

\FIGURE{
\centering
  \includegraphics[width=0.49\textwidth]{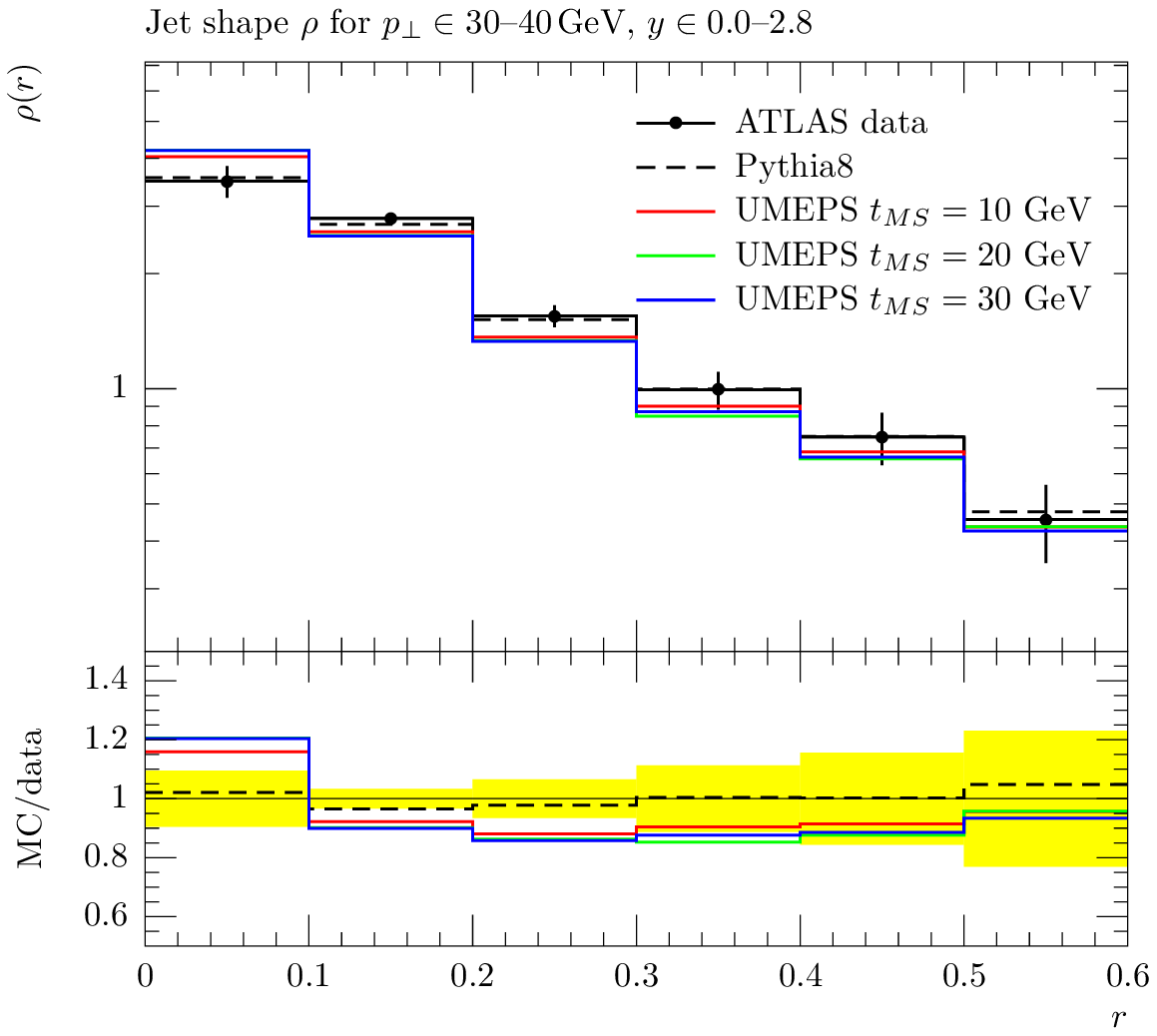}
  \includegraphics[width=0.49\textwidth]{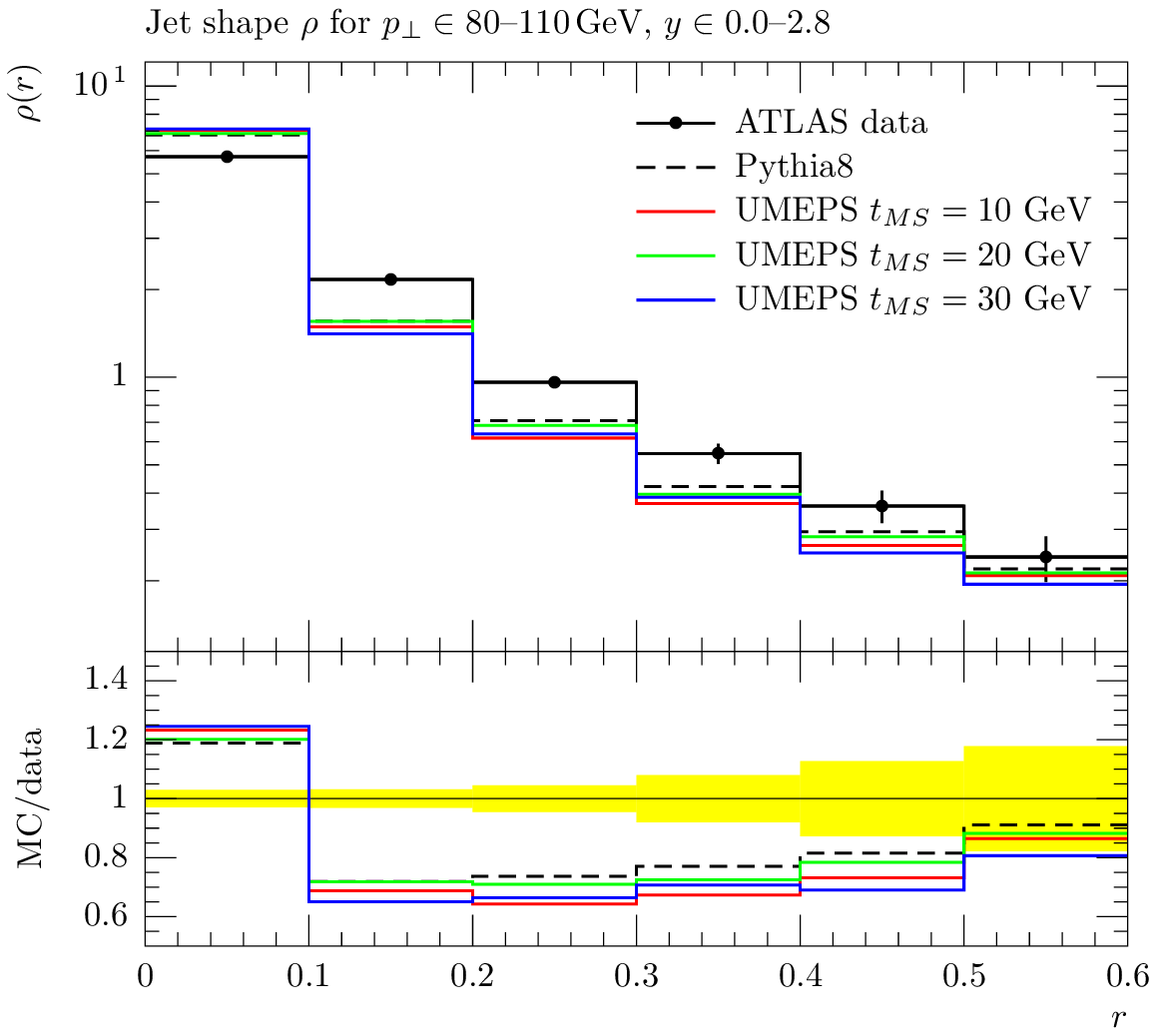}
  \caption{\label{fig:atlas-qcd-rho}Jet shapes in QCD events, for three 
  different merging scales, in two $p_\perp$ bins, as measured by ATLAS 
  \cite{Aad:2011kq}. Effects of multiple scatterings and hadronisation are 
  included.}
}

\FIGURE{
\centering
  \includegraphics[width=0.49\textwidth]{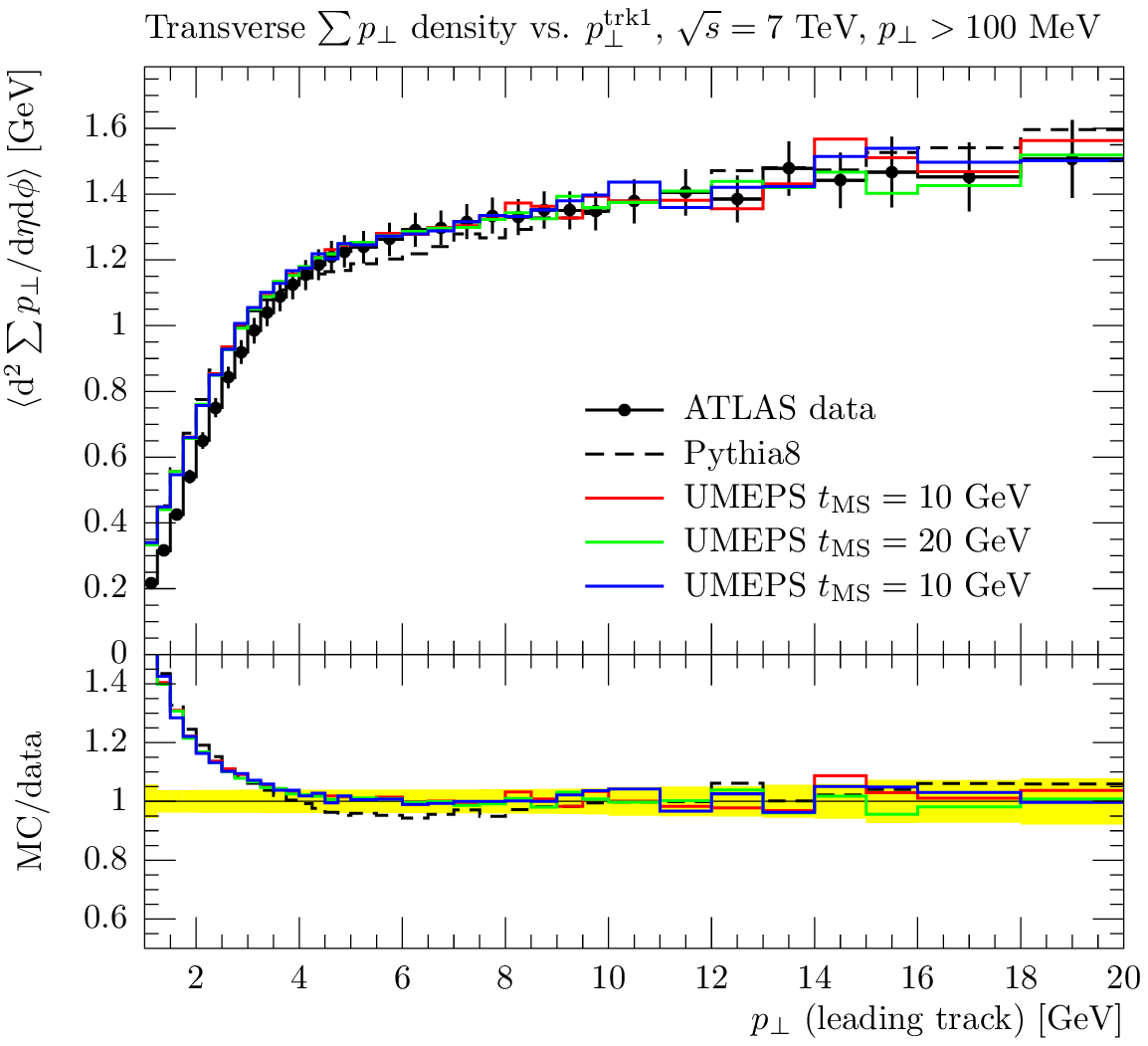}
  \includegraphics[width=0.49\textwidth]{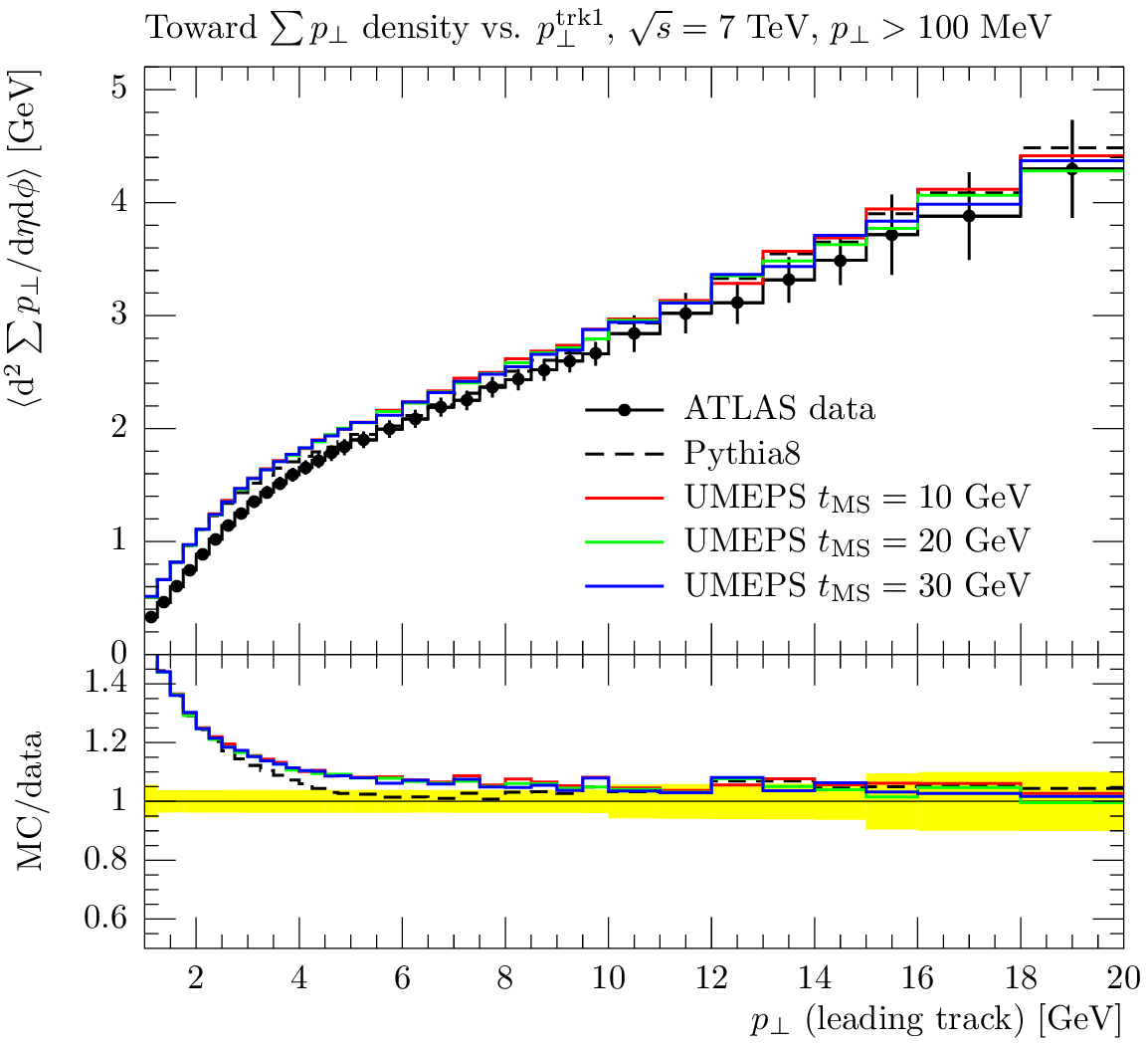}
  \caption{\label{fig:atlas-qcd-ptsum}Sum of transverse momenta of charged
  particles in QCD events, for three different merging scales, in the
  transverse and toward region, as measured by ATLAS \cite{Aad:2010fh}.
  Effects of multiple scatterings and hadronisation are 
  included.}
}

Before moving to the discussion section, we would like to investigate
how jet shapes at the LHC are changed by the inclusion of additional
matrix elements in the pure QCD case. The ATLAS analysis
\cite{Aad:2011kq} found that the differential jet shape for relatively
low-$p_\perp$ jets with $p_\perp \leq 160$ GeV depends crucially on
the modelling of the underlying event. In Figure
\ref{fig:atlas-qcd-rho}, we show the default \pytppp and UMEPS results
for two $p_\perp$ bins. Although far from perfect,
the difference between the pure shower and merged results are similar
to what a minor change in \as\ would give\footnote{This of course
does \emph{not} mean that we will to use this data for tuning, but that
$\as$-choices for different tunes can have a comparable effect on $\rho(r)$.}. 
We are confident that
the prescription for adding MPI (see section
\ref{sec:umeps-step-by-step}) does indeed mean that the
underlying-event modelling of \pytppp is only marginally perturbed by
the inclusion of additional jets.
This is supported by Figure \ref{fig:atlas-qcd-ptsum},
which shows the sum of charged-particle transverse momenta in region
close to the leading track (i.e.\ the toward region) and perpendicular
to the leading track (transverse region) \cite{Aad:2010fh}. These are
typical minimum-bias observables especially designed to investigate
the underlying event, and \pytppp tune A2 includes this data in the
tuning procedure. It is reassuring that the inclusion of two
additional jets through the UMEPS scheme did not invalidate this tuning.

\section{Discussion}\label{sec:discussion}

Before concluding this letter, we would like to make some comments to put 
this method into perspective.

\subsubsection*{Relation to LoopSim}

Even though not completely obvious at first sight, UMEPS was heavily 
influenced by the LoopSim method \cite{Rubin:2010xp}. This method as introduced to tame 
order-by-order large logarithmic enhancements by combining matrix elements
with different jet multiplicities in a unitary way. The combination is done
by joining all combinations to integrate over $1,\dots,n$ partons in the 
ME event $\state{n}$, and also allowing an integration over hard process
particles. In figure~\ref{fig:incomplete}, the final state gluons are 
candidates for integration (or, in the terminology of LoopSim, looping), and
the $\W-$boson can be looped as well. With such a procedure, enhancements due
to collinear $\W-$boson radiation off a dijet state can be 
compensated\footnote{At very large transverse momenta, the effect of 
multiple soft/collinear electroweak bosons becomes important. An
appraisal of high-$p_\perp$ observables in $\W\Z$-production has recently been
reported \cite{Campanario:2012fk}.}.

Apart from major technical differences, one interesting difference is that 
in the LoopSim method, higher fixed-order corrections are approximated by 
multiple loopings, whereas in UMEPS, an all-order expression is included in 
the $\state{n+1}$ state before one parton is looped. Integrating multiple
emissions is only necessary for $\ordms-$unordered sequences of splittings, 
which are not considered in LoopSim. Furthermore, in UMEPS, only integrations
of QCD splittings are performed, while LoopSim includes loopings of 
$\W-$boson radiation. We postpone the inclusion $\W-$boson clusterings in 
UMEPS until a full electroweak shower is available in \pytppp. It would 
clearly be interesting to combine the Sudakov resummation in UMEPS with the 
multiple loopings done in LoopSim. To arrive at a better description of 
$\state{0}$ configurations, one could e.g.\ take $\as^{1}$ contributions from
the looped $\state{1}$ state, $\as^{2}$ contributions from
the twice-looped $\state{2}$, and all higher orders from
Sudakov-reweighted thrice-integrated $\state{3}$ events. The way the inclusive
cross section is maintained in such a procedure will be less obvious than in 
the case of LoopSim or UMEPS.

\subsubsection*{Merging scale dependence}

In the original CKKW-L algorithm it is evident that the dependence on
the merging scale is absent to the accuracy of the PS. This means that
for any observable, leading logarithmic terms on the form
$\as^nL^{2n}$, where $L=\ln\muf/\ordms$, are correctly cancelled to all
orders. For a shower\footnote{Note that the shower in \pytppp which is
  used in simulations in this paper has not been formally proven to be
  NLL-correct.} which in addition is correct to next-to-leading
logarithmic accuracy, also terms on the form $\as^nL^{2n-1}$ are
correctly cancelled.

What we have accomplished with the UMEPS method is that the total
inclusive Born-level cross section is almost completely independent of
the merging scale. In addition, if we look at the master formula in
\eqref{eq:UMEPS-master}, it is clear that for any phase space point
$\phi_n$ with $n$ resolved partons, the inclusive cross section,
integrating all contributions from higher parton multiplicities, is
simply given by $\Tev{n}$. Although $\Tev{n}$ includes no-emission
probabilities calculated by the shower, it has no dependence on the
merging scale, and hence, all inclusive $n$-jet cross sections are
independent of the merging scale. Since exclusive $n$-jet cross
sections are the difference between the $n$-jet and $n+1$-jet
inclusive cross sections, also these are independent of the merging
scale.

The independence is, of course, not exact for any real observable. A
jet algorithm will not cluster an $n+1$-jet state back to the precise
$n$-parton phase space point as would the mapping of the parton
shower, or symbolically for a general observable $\mathcal{O}$,
$\int\mathcal{O}({\state{n+1j}})\Tev{n+1}\ne
\mathcal{O}({\state{nj}})\Iev{n+1}{n}$. However, as long as the
observable is collinear- and infrared-safe this difference will not
have any logarithmic enhancements, and as long as the $n$-jet state is
well above the merging scale we can take this scale to be arbitrarily
small, without changing the $n$-jet cross section.

Also, as we have noted before, there are some $n$-parton states, such
as the one in figure~\ref{fig:incomplete}, which do not have an
underlying $n-1$-parton state reachable with a reconstructed PS
emission. Unless the PS is amended with $W$-strahlung splittings,
such contributions will always give a small dependence. However, we
have found these to be numerically very small in the cases we have
investigated.

\subsubsection*{Events with negative weight}

Contrary to the standard CKKW(-L) algorithms our new UMEPS procedure
will produce negatively weighted events. There has in the past been a
great reluctance in the experimental community towards using
generators with negative event weights. Mostly this has been a
question about problems in handling the statistics and that it seems
wasteful to spend a huge number of CPU cycles to do a full detector
simulation on an event, which in the end will be cancelled by another
event with a negative weight. However, the acceptance for negative
weights have increased, and today most experiments are using programs
such as MC@NLO
\cite{Frixione:2006gn,Hoeche:2011fd,Hoeche:2012ft,Hirschi:2011pa},
which do produce a fair amount of events with negative weight.

Clearly UMEPS is more wasteful than CKKW-L in this respect, and the
number of events that need to be analysed to get the same statistics
is more than doubled. In fact it can be shown that the variance in the
event weights, when calculating the no-emission probability for the
zero-jet case in CKKW-L with the Sudakov-veto algorithm, is proportional
to $\noem{0}-\noem{0}^2$. The variance for UMEPS, where the
corresponding factor is calculated by reclustering one-jet states
multiplied by a no-emission probability, is of the form
\begin{equation}
  \frac{1-\noem{0}}{-\ln\noem{0}}-\left(\frac{1-\noem{0}}{-\ln\noem{0}}\right)^2.
  \label{eq:umeps-variance}
\end{equation}
Thus, for small merging scales (giving small no-emission probabilities),
UMEPS becomes very inefficient as compared to CKKW-L.

We believe that the benefits of UMEPS outweigh this drawback. Also we
note that the algorithm works in a way such that all events will
either have zero weight or a weight of order ($\pm$) unity. This is
because the no-emission probabilities are generated by the
Sudakov-veto algorithm and are therefore either zero or unity, while
the PDF- and \as-reweighting typically is of ${\cal O}(1)$. Had the
no-emission probabilities been calculated analytically, they would be
very small for small merging scales, and \eg each single 0-jet ME
event would have to be cancelled by large number of small-weight
reclustered 1-jet events. This would be very inefficient if a
CPU-heavy detector simulation would have to be run on each event.

On the other hand, the Sudakov-veto algorithm causes some problem in
the case full detector simulation is not used. In this case the
computational bottle neck is typically in the ME generation of high
jet multiplicities, and the problem is that most of these events are
given zero weight and will be thrown away by the Sudakov-veto
algorithm, especially for small merging scales. This can in principle
be handled by a modification of the veto algorithm
\cite{lonnblad:2012..} where all events are kept but are given a small
weight.

\section{Conclusions and Outlook}
\label{sec:conclusions-outlook}

In this article, we have presented a new method for tree-level matrix element
merging called UMEPS. This method is heavily indebted to the CKKW-L 
prescription, but explicitly keeps the total inclusive cross section fixed. 
Since it builds on the implementation of CKKW-L in \pytppp, all developments 
of CKKW-L are immediately available to UMEPS. This for example includes 
improvements for BSM processes \cite{Dreiner:2012gx} and multiple pre-defined
merging scales.

The UMEPS scheme uses an add-subtract prescription inspired by parton shower 
unitarity to combine the improved description of observable shapes of CKKW-L 
with a fixed total inclusive cross section. This means that significantly 
lower the merging scale values are possible, which allows for controlled 
improvement of low-scale features of the parton shower. Tuning efforts will 
be subject of a future article. When confronted with data, UMEPS and CKKW-L
perform equally well.

UMEPS is an ideal candidate for further improvements, since the 
lowest-multiplicity cross section is not reweighted, making replacements 
with the full NLO or NNLO results possible. We have successfully implemented
an NLO extension, and will present it in a separate publication \cite{prestel:2012..}.

Finally, while finishing this article, it came to our
attention that a very similar approach has been developed in parallel 
by Pl\"{a}tzer \cite{Platzer:2012bs}, which further describes the extension of 
an inclusive-cross-section preserving merging method to NLO accuracy.

\section*{Acknowledgements}

Work supported in part by the Swedish research council (contracts
621-2009-4076 and 621-2010-3326). We would like to thank Simon Pl\"{a}tzer,
Stefan H\"{o}che and Frank Tackmann for helpful discussions.

\bibliographystyle{utcaps}  
\bibliography{references} 

\providecommand{\href}[2]{#2}\begingroup\raggedright\begin{thebibliography}{10}

\bibitem{Lavesson:2008ah}
N.~Lavesson and L.~Lönnblad,
  \href{http://dx.doi.org/10.1088/1126-6708/2008/12/070}{{\em JHEP} {\bf 12}
  (2008)  070},
\href{http://arxiv.org/abs/0811.2912}{{\tt arXiv:0811.2912 [hep-ph]}}.

\bibitem{Hamilton:2010wh}
K.~Hamilton and P.~Nason, \href{http://dx.doi.org/10.1007/JHEP06(2010)039}{{\em
  JHEP} {\bf 1006} (2010)  039},
\href{http://arxiv.org/abs/1004.1764}{{\tt arXiv:1004.1764 [hep-ph]}}.

\bibitem{Hoche:2010kg}
S.~H{\"o}che, F.~Krauss, M.~Sch{\"o}nherr, and F.~Siegert,
\href{http://arxiv.org/abs/1009.1127}{{\tt arXiv:1009.1127 [hep-ph]}}.

\bibitem{Alioli:2011nr}
S.~Alioli, K.~Hamilton, and E.~Re,
  \href{http://dx.doi.org/10.1007/JHEP09(2011)104}{{\em JHEP} {\bf 1109} (2011)
   104},
\href{http://arxiv.org/abs/1108.0909}{{\tt arXiv:1108.0909 [hep-ph]}}.

\bibitem{Hamilton:2012np}
K.~Hamilton, P.~Nason, and G.~Zanderighi,
  \href{http://dx.doi.org/10.1007/JHEP10(2012)155}{{\em JHEP} {\bf 1210} (2012)
   155},
\href{http://arxiv.org/abs/1206.3572}{{\tt arXiv:1206.3572 [hep-ph]}}.

\bibitem{Gehrmann:2012yg}
T.~Gehrmann, S.~H{\"o}che, F.~Krauss, M.~Sch{\"o}nherr, and F.~Siegert,
\href{http://arxiv.org/abs/1207.5031}{{\tt arXiv:1207.5031 [hep-ph]}}.

\bibitem{Hoeche:2012yf}
S.~H{\"o}che, F.~Krauss, M.~Sch{\"o}nherr, and F.~Siegert,
\href{http://arxiv.org/abs/1207.5030}{{\tt arXiv:1207.5030 [hep-ph]}}.

\bibitem{Frederix:2012ps}
R.~Frederix and S.~Frixione,
\href{http://arxiv.org/abs/1209.6215}{{\tt arXiv:1209.6215 [hep-ph]}}.

\bibitem{Bauer:2008qh}
C.~W. Bauer, F.~J. Tackmann, and J.~Thaler,
  \href{http://dx.doi.org/10.1088/1126-6708/2008/12/010}{{\em JHEP} {\bf 0812}
  (2008)  010},
\href{http://arxiv.org/abs/0801.4026}{{\tt arXiv:0801.4026 [hep-ph]}}.

\bibitem{Bauer:2008qj}
C.~W. Bauer, F.~J. Tackmann, and J.~Thaler,
  \href{http://dx.doi.org/10.1088/1126-6708/2008/12/011}{{\em JHEP} {\bf 0812}
  (2008)  011},
\href{http://arxiv.org/abs/0801.4028}{{\tt arXiv:0801.4028 [hep-ph]}}.

\bibitem{scet}
C.~Bauer {\em et al.}, {\em {(preprint in preparation)}}  . See eg.\ talk
  presented at the PSR12 workshop
  \href{https://indico.desy.de/getFile.py/access?contribId=21&resId=0&materialId=slides&confId=5230}{\texttt{https://indico.desy.de/conferenceTimeTable.py?confId=5230\#20120530}}.

\bibitem{Catani:2001cc}
S.~Catani, F.~Krauss, R.~Kuhn, and B.~R. Webber, {\em JHEP} {\bf 11} (2001)
  063,
\href{http://arxiv.org/abs/hep-ph/0109231}{{\tt arXiv:hep-ph/0109231}}.

\bibitem{Lonnblad:2001iq}
L.~Lönnblad, {\em JHEP} {\bf 05} (2002)  046,
\href{http://arxiv.org/abs/hep-ph/0112284}{{\tt arXiv:hep-ph/0112284}}.

\bibitem{Lavesson:2005xu}
N.~Lavesson and L.~Lönnblad, {\em JHEP} {\bf 07} (2005)  054,
\href{http://arxiv.org/abs/hep-ph/0503293}{{\tt arXiv:hep-ph/0503293}}.

\bibitem{Hoeche:2009rj}
S.~H{\"o}che, F.~Krauss, S.~Schumann, and F.~Siegert,
  \href{http://dx.doi.org/10.1088/1126-6708/2009/05/053}{{\em JHEP} {\bf 05}
  (2009)  053},
\href{http://arxiv.org/abs/0903.1219}{{\tt arXiv:0903.1219 [hep-ph]}}.

\bibitem{Lonnblad:2011xx}
{L.~L\"onnblad, and S.~Prestel},
  \href{http://dx.doi.org/10.1007/JHEP03(2012)019}{{\em JHEP} {\bf 1203} (2012)
   019},
\href{http://arxiv.org/abs/1109.4829}{{\tt arXiv:1109.4829 [hep-ph]}}.

\bibitem{Bengtsson:1986et}
M.~Bengtsson and T.~Sjostrand,
\href{http://dx.doi.org/10.1016/0550-3213(87)90407-X}{{\em Nucl.Phys.} {\bf
  B289} (1987)  810}.

\bibitem{Bengtsson:1986hr}
M.~Bengtsson and T.~Sjostrand,
\href{http://dx.doi.org/10.1016/0370-2693(87)91031-8}{{\em Phys.Lett.} {\bf
  B185} (1987)  435}.

\bibitem{Miu:1998ut}
G.~Miu,
\href{http://arxiv.org/abs/hep-ph/9804317}{{\tt arXiv:hep-ph/9804317
  [hep-ph]}}.

\bibitem{Miu:1998ju}
G.~Miu and T.~Sjostrand,
  \href{http://dx.doi.org/10.1016/S0370-2693(99)00068-4}{{\em Phys.Lett.} {\bf
  B449} (1999)  313--320},
\href{http://arxiv.org/abs/hep-ph/9812455}{{\tt arXiv:hep-ph/9812455
  [hep-ph]}}.

\bibitem{Giele:2007di}
W.~T. Giele, D.~A. Kosower, and P.~Z. Skands,
  \href{http://dx.doi.org/10.1103/PhysRevD.78.014026}{{\em Phys.Rev.} {\bf D78}
  (2008)  014026},
\href{http://arxiv.org/abs/0707.3652}{{\tt arXiv:0707.3652 [hep-ph]}}.

\bibitem{Giele:2011cb}
W.~Giele, D.~Kosower, and P.~Skands,
  \href{http://dx.doi.org/10.1103/PhysRevD.84.054003}{{\em Phys.Rev.} {\bf D84}
  (2011)  054003},
\href{http://arxiv.org/abs/1102.2126}{{\tt arXiv:1102.2126 [hep-ph]}}.

\bibitem{Sjostrand:2007gs}
T.~Sjöstrand, S.~Mrenna, and P.~Skands,
  \href{http://dx.doi.org/10.1016/j.cpc.2008.01.036}{{\em Comput. Phys.
  Commun.} {\bf 178} (2008)  852--867},
\href{http://arxiv.org/abs/0710.3820}{{\tt arXiv:0710.3820 [hep-ph]}}.

\bibitem{Nason:2004rx}
P.~Nason, \href{http://dx.doi.org/10.1088/1126-6708/2004/11/040}{{\em JHEP}
  {\bf 11} (2004)  040},
\href{http://arxiv.org/abs/hep-ph/0409146}{{\tt arXiv:hep-ph/0409146}}.

\bibitem{Alwall:2006yp}
J.~Alwall {\em et al.}, {\em Comput. Phys. Commun.} {\bf 176} (2007)  300--304,
  \href{http://arxiv.org/abs/hep-ph/0609017}{{\tt hep-ph/0609017}}.

\bibitem{Cacciari:2011ma}
M.~Cacciari, G.~P. Salam, and G.~Soyez,
  \href{http://dx.doi.org/10.1140/epjc/s10052-012-1896-2}{{\em Eur.Phys.J.}
  {\bf C72} (2012)  1896},
\href{http://arxiv.org/abs/1111.6097}{{\tt arXiv:1111.6097 [hep-ph]}}.

\bibitem{Balazs:2000sz}
C.~Balazs, J.~Huston, and I.~Puljak,
  \href{http://dx.doi.org/10.1103/PhysRevD.63.014021}{{\em Phys.Rev.} {\bf D63}
  (2001)  014021},
\href{http://arxiv.org/abs/hep-ph/0002032}{{\tt arXiv:hep-ph/0002032
  [hep-ph]}}.

\bibitem{Corke:2010yf}
R.~Corke and T.~Sjöstrand,
  \href{http://dx.doi.org/10.1007/JHEP03(2011)032}{{\em JHEP} {\bf 03} (2011)
  032},
\href{http://arxiv.org/abs/1011.1759}{{\tt arXiv:1011.1759 [hep-ph]}}.

\bibitem{ATL:2012-003pub}

T.~A. collaboration.

\bibitem{Buckley:2010ar}
A.~Buckley {\em et al.},
\href{http://arxiv.org/abs/1003.0694}{{\tt arXiv:1003.0694 [hep-ph]}}.

\bibitem{Aad:2012en}
{\bf ATLAS Collaboration} Collaboration, G.~Aad {\em et al.},
  \href{http://dx.doi.org/10.1103/PhysRevD.85.092002}{{\em Phys.Rev.} {\bf D85}
  (2012)  092002},
\href{http://arxiv.org/abs/1201.1276}{{\tt arXiv:1201.1276 [hep-ex]}}.

\bibitem{Aad:2011kq}
{\bf Atlas Collaboration} Collaboration, G.~Aad {\em et al.},
  \href{http://dx.doi.org/10.1103/PhysRevD.83.052003}{{\em Phys.Rev.} {\bf D83}
  (2011)  052003},
\href{http://arxiv.org/abs/1101.0070}{{\tt arXiv:1101.0070 [hep-ex]}}.

\bibitem{Aad:2010fh}
{\bf Atlas Collaboration} Collaboration, G.~Aad {\em et al.},
  \href{http://dx.doi.org/10.1103/PhysRevD.83.112001}{{\em Phys.Rev.} {\bf D83}
  (2011)  112001},
\href{http://arxiv.org/abs/1012.0791}{{\tt arXiv:1012.0791 [hep-ex]}}.

\bibitem{Rubin:2010xp}
M.~Rubin, G.~P. Salam, and S.~Sapeta,
  \href{http://dx.doi.org/10.1007/JHEP09(2010)084}{{\em JHEP} {\bf 1009} (2010)
   084},
\href{http://arxiv.org/abs/1006.2144}{{\tt arXiv:1006.2144 [hep-ph]}}.

\bibitem{Campanario:2012fk}
F.~Campanario and S.~Sapeta,
  \href{http://dx.doi.org/10.1016/j.physletb.2012.10.013}{{\em Phys.Lett.} {\bf
  B718} (2012)  100--104},
\href{http://arxiv.org/abs/1209.4595}{{\tt arXiv:1209.4595 [hep-ph]}}.

\bibitem{Frixione:2006gn}
S.~Frixione and B.~R. Webber,
\href{http://arxiv.org/abs/hep-ph/0612272}{{\tt arXiv:hep-ph/0612272}}.

\bibitem{Hoeche:2011fd}
S.~H{\"o}che, F.~Krauss, M.~Sch{\"o}nherr, and F.~Siegert,
  \href{http://dx.doi.org/10.1007/JHEP09(2012)049}{{\em JHEP} {\bf 1209} (2012)
   049},
\href{http://arxiv.org/abs/1111.1220}{{\tt arXiv:1111.1220 [hep-ph]}}.

\bibitem{Hoeche:2012ft}
S.~H{\"o}che, F.~Krauss, M.~Sch{\"o}nherr, and F.~Siegert, {\em Physical Review
  Letters} (2012)  ,
\href{http://arxiv.org/abs/1201.5882}{{\tt arXiv:1201.5882 [hep-ph]}}.

\bibitem{Hirschi:2011pa}
V.~Hirschi, R.~Frederix, S.~Frixione, M.~V. Garzelli, F.~Maltoni, {\em et al.},
  \href{http://dx.doi.org/10.1007/JHEP05(2011)044}{{\em JHEP} {\bf 1105} (2011)
   044},
\href{http://arxiv.org/abs/1103.0621}{{\tt arXiv:1103.0621 [hep-ph]}}.

\bibitem{lonnblad:2012..}
L.~L{\"o}nnblad,
\href{http://arxiv.org/abs/1211.7204}{{\tt arXiv:1211.7204 [hep-ph]}}.

\bibitem{Dreiner:2012gx}
H.~K. Dreiner, M.~Kramer, and J.~Tattersall,
  \href{http://dx.doi.org/10.1209/0295-5075/99/61001}{{\em Europhys.Lett.} {\bf
  99} (2012)  61001},
\href{http://arxiv.org/abs/1207.1613}{{\tt arXiv:1207.1613 [hep-ph]}}.

\bibitem{prestel:2012..}
L.~L{\"o}nnblad and S.~Prestel,
\href{http://arxiv.org/abs/1211.7278}{{\tt arXiv:1211.7278 [hep-ph]}}.

\bibitem{Platzer:2012bs}
S.~Pl{\"a}tzer,
\href{http://arxiv.org/abs/1211.5467}{{\tt arXiv:1211.5467 [hep-ph]}}.

\end{thebibliography}\endgroup

\end{document}